\documentclass[Journal]{IEEEtran}
\IEEEoverridecommandlockouts
\usepackage{cite}
\usepackage{amsmath,amssymb,amsfonts}

\usepackage{textcomp}
\usepackage{xcolor}
\usepackage{multirow}
\usepackage{makecell}
\usepackage{algorithm,algcompatible}
\usepackage{subfigure}
\usepackage{graphicx}
\usepackage{array}
\newcolumntype{P}[1]{>{\centering\arraybackslash}p{#1}}

\usepackage{booktabs}
\newtheorem{lemma}{Lemma}
\newtheorem{theorem}{Theorem}

\def\BibTeX{{\rm B\kern-.05em{\sc i\kern-.025em b}\kern-.08em
    T\kern-.1667em\lower.7ex\hbox{E}\kern-.125emX}}

\long\def\comment#1{}

\newfont{\bbb}{msbm10 scaled 700}

\newfont{\bb}{msbm10 scaled 1100}
\newcommand{\CC}{\mbox{\bb C}}

\newcommand{\EE}{\mbox{\bb E}}

\newcommand{\av}{{\bf a}}
\newcommand{\bv}{{\bf b}}
\newcommand{\cv}{{\bf c}}
\newcommand{\dv}{{\bf d}}

\newcommand{\fv}{{\bf f}}

\newcommand{\hv}{{\bf h}}

\newcommand{\nv}{{\bf n}}

\newcommand{\tv}{{\bf t}}
\newcommand{\uv}{{\bf u}}
\newcommand{\wv}{{\bf w}}
\newcommand{\vv}{{\bf v}}
\newcommand{\xv}{{\bf x}}
\newcommand{\yv}{{\bf y}}
\newcommand{\zv}{{\bf z}}
\newcommand{\zerov}{{\bf 0}}

\newcommand{\Am}{{\bf A}}

\newcommand{\Cm}{{\bf C}}
\newcommand{\Dm}{{\bf D}}

\newcommand{\Fm}{{\bf F}}
\newcommand{\Gm}{{\bf G}}
\newcommand{\Hm}{{\bf H}}
\newcommand{\Id}{{\bf I}}

\newcommand{\Mm}{{\bf M}}
\newcommand{\Nm}{{\bf N}}

\newcommand{\Sm}{{\bf S}}
\newcommand{\Tm}{{\bf T}}
\newcommand{\Um}{{\bf U}}
\newcommand{\Wm}{{\bf W}}
\newcommand{\Vm}{{\bf V}}
\newcommand{\Xm}{{\bf X}}
\newcommand{\Ym}{{\bf Y}}
\newcommand{\Zm}{{\bf Z}}

\newcommand{\Mc}{{\cal M}}

\newcommand{\Oc}{{\cal O}}
\newcommand{\Pc}{{\cal P}}

\newcommand{\Vc}{{\cal V}}

\newcommand{\nuv}{\hbox{\boldmath$\nu$}}

\newcommand{\Lambdam}{\hbox{\boldmath$\Lambda$}}
\newcommand{\Deltam}{\hbox{\boldmath$\Delta$}}

\newcommand{\Phim}{\hbox{\boldmath$\Phi$}}

\newcommand{\SNR}{{\sf SNR}}

\newcommand{\eqdef}{\stackrel{\Delta}{=}}

\newcommand{\herm}{{\sf H}}

\newcommand{\RED}{\color[rgb]{1.00,0.10,0.10}}
\newcommand{\BLUE}{\color[rgb]{0,0,0.90}}

\DeclareMathOperator*{\argmax}{arg\,max}
\DeclareMathOperator*{\argmin}{arg\,min}

\newtheorem{remark}{Remark}

\begin{document}

\title{Piecewise Beam Training and Channel Estimation for RIS-Aided Near-Field Communications}
\author{Jeongjae~Lee,~\IEEEmembership{Student Member,~IEEE}        and~Songnam~Hong,~\IEEEmembership{Member,~IEEE}
\thanks{J. Lee and S. Hong are with the Department of Electronic Engineering, Hanyang University, Seoul, Korea (e-mail: \{lyjcje7466, snhong\}@hanyang.ac.kr).}

\thanks{This work was supported in part by the Institute of Information \& communications Technology Planning \& Evaluation (IITP) under the artificial intelligence semiconductor support program to nurture the best talents (IITP-2025-RS-2023-00253914) grant funded by the Korea government(MSIT) and in part by the National Research Foundation of Korea(NRF) grant funded by the Korea government(MSIT)(No. RS-2024-00409492).}
}

\maketitle

\begin{abstract}
In this paper, we investigate the channel estimation challenge in reconfigurable intelligent surface (RIS)-aided near-field communication systems. Current channel estimation techniques require substantial pilot overhead and computational complexity, especially when the number of RIS elements is extremely large. To address this issue, we introduce a two-timescale channel estimation strategy that leverages the asymmetric coherence times of both the RIS-base station (BS) channel and the User-RIS channel. We derive a time-scaling property indicating that, for any two effective channels within the longer coherence time, one effective channel can be represented as the product of a vector, termed the small-timescale effective channel, and the other effective channel. By integrating the estimated effective channel from the initial time block with observations from our piecewise beam training, we present an efficient method for estimating subsequent small-timescale effective channels. We theoretically verify the efficacy of the proposed RIS design and demonstrate, through simulations, that our channel estimation method outperforms existing methods in pilot overhead and computational complexity across various realistic channel models.


\end{abstract}



\section{Introduction}\label{sec:intro}
With the growth of various emerging applications such as autonomous vehicles, virtual reality, and holograms, future wireless communication systems aim to support ultra-high data rates \cite{dang2020should,Cha2023,Wang2024}. To this end, it is imperative to develop an efficient technique to mitigate severe pathloss in high-frequency communications, such as millimeter wave (mmWave) and terahertz (THz) \cite{wang2018millimeter,Jiang2024}. The extremely large-scale antenna array (ELAA) has been proposed to substantially enhance beamforming gain in massive multiple-input multiple-output (MIMO) systems, thereby effectively mitigating loss \cite{Carvalho2020}. In MIMO systems with ELAA, referred to as XL-MIMO, hundreds to thousands of antennas form a uniform antenna array, enabling specific beam patterns to be generated. Additionally, a reconfigurable intelligence surface (RIS), positioned between a base station (BS) and users, can physically enhance blocked or weakened channels by favorably manipulating incident waves, thereby improving overall communication performance \cite{di2020smart, pei2021ris, wu2021intelligent}. Owing to its lower implementation costs and energy loss, this technology is considered for overcoming the physical limitations of high-frequency communications by establishing new and strengthened channels.

By leveraging the advantages of both techniques, the integration of ELAA and RIS has been recently explored in the current literature \cite{Mu2024,Lv2024}. In contrast to conventional RIS-aided communication systems, such as RIS-aided massive MIMO systems \cite{Zhou2024}, the spherical near-field effect must be considered due to the electromagnetic characteristic of ELAA \cite{Zhou2015}. This integrated system is designated as the RIS-aided {\em near-field} communication system and exhibits characteristics akin to a double-edged sword. Enhanced by a double-sided line-of-sight (LoS) channel and precise beamfocusing capability, there exists a novel opportunity in RIS-aided near-field communication systems that improves the degree of freedom (DoF) \cite{Mu2024}. Nevertheless, estimating the channels or optimizing beam focusing becomes extremely challenging due to the significantly increased number of parameters associated with higher rank \cite{Lu2023}, as well as the notable emergence of substantial beam splitting (or beam squint) in wideband systems \cite{Cui2024,Cui2023rainbow}. We in this paper contribute to the channel estimation problem in RIS-aided near-field communication systems.

\subsection{Related Works}\label{subsec;relatedwork}
Over the past five years, the channel estimation problem for the RIS-aided MIMO systems has been widely investigated \cite{chen2023channel,schroeder2022channel,Yang2023,Yang2024,Lee2024near,Lee2025}. The objective is to acquire the channel between the RIS and the base station (BS), referred to as the RIS-BS channel, as well as the channels between the users and the RIS, termed the User-RIS channels. However, estimating these channels separately is impractical due to the lack of a signal processing unit at the passive RIS. Considering both practicality and usefulness, existing works have concentrated on estimating the effective (or cascaded) channels while maintaining an affordable pilot overhead. To develop an efficient channel estimation method, one representative approach is to leverage the sparsity of mmWave or THz channels. These sparsity-based methods, such as compressed sensing (CS) \cite{chen2023channel} and atomic norm minimization (ANM) \cite{schroeder2022channel}, demonstrate significant performance improvements for the {\em far-field} effective channels, where array response vectors are linear with respect to channel parameters such as the angle-of-arrival (AoA) and angle-of-departure (AoD) in the planar wave approximation.

Despite the channel sparsity, employing the prior works in RIS-aided near-field MIMO systems presents significant challenges. This difficulty arises from the non-linearity of the near-field array response vector, which is a consequence of its spherical wavefronts. In \cite{Yang2023,Yang2024}, this challenging problem was examined under an ideal scenario: the LoS far-field RIS-BS channel is known a priori, while the near-field User-RIS channels are solely estimated. In this scenario, a CS-based channel estimation method was proposed for the RIS-aided near-field MIMO systems, leveraging the polar-domain sparsity introduced in \cite{Lu2023}. Notably, the dictionary matrix differs significantly from those used in CS-based far-field counterparts \cite{chen2023channel,schroeder2022channel}, as it must be carefully designed according to the structures of the array response vectors. Recently in \cite{Lee2024near}, an efficient channel estimation method, referred to as CLRA, was proposed for RIS-aided near-field MIMO systems (also known as XL-RIS assisted XL-MIMO systems). This method was developed by leveraging the low-rankness of the RIS-BS channel. Remarkably, the CLRA can be seamlessly applied to both far-field and near-field channels without any modification, whereas CS-based methods necessitate a suitable design of the dictionary matrix tailored to the categories of wireless channels. Furthermore, it has been demonstrated that the CLRA achieves higher estimation accuracy compared to the corresponding CS-based methods while maintaining lower training overhead.

Unfortunately, the assumption of channel sparsity or low-rankness for the effective channel becomes invalid when the distance between the RIS and the BS falls below a threshold known as MIMO advanced Rayleigh distance (MIMO-ARD) \cite{Lu2023}. In this scenario, the double-sided LoS channel matrix for the RIS-BS link necessitates careful modeling that considers more precise channel elements. As a result, the effective channel exhibits a high rank, losing its sparsity. This situation facilitates beamfocusing for multi-user or multi-stream communications but poses significant challenges for channel estimation.
Very recently, \cite{Lee2025} introduced a channel estimation method called PW-CLRA, which partitions the RIS into multiple subarrays to effectively manage the high rank of the effective channel, achieving notable estimation accuracy with low overhead. However, the required pilot overhead remains substantial, particularly as the number of RIS elements increases. 

In \cite{Hu2021}, a two-timescale channel estimation strategy, referred to as {\bf 2TCE}, was introduced to reduce pilot overhead by exploiting the asymmetry between the coherence times of the RIS-BS and User-RIS channels. While this method demonstrates notable performance under the Rayleigh fading channel, it does not ensure optimal estimation performance for channels in mmWave or THz communications.  Furthermore, it is limited to {\em full-duplex BS operation}, which restricts its applicability in various RIS-aided communication systems, and it lacks a theoretical analysis for the design of RIS reflection vector. Nonetheless, there remains potential to adapt this 2TCE strategy for high-rank near-field effective channels by addressing these limitations, which is the motivation of this paper.

\subsection{Our Contributions}
This paper investigates the channel estimation problem for RIS-aided near-field communication systems. Considering a realistic near-field channel model, especially, we explore a channel estimation method tailored for the 2TCE framework, particularly applicable when the BS operates in practical half-duplex mode and utilizes a hybrid beamforming architecture. This framework comprises two phases: i) large-timescale channel estimation, conducted once within the coherence time of the RIS-BS channel; ii) small-timescale channel estimation, performed multiple times. We propose a novel small-scale channel estimation method along with the associated beam training method to enhance accuracy while reducing training overhead and computational complexity. Our key contributions are summarized as follows:
\begin{itemize}
  
    \item We consider a realistic near-field scenario in mmWave communications, presenting an accurate model of the double-sided LoS MIMO channel between the BS and the RIS, which is influenced by the substantial number of antennas at both the BS and RIS. Additionally, we account for the blocking effect on the RIS, which reflects the visual region (VR). Based on this framework, we define the effective channel to be estimated in this paper and analyze its rank in conjunction with the existing sparse and Rayleigh fading channels to formulate an appropriate channel estimation problem. 
    
    \item We derive a time-scaling property indicating that for any two effective channels within the coherence time of the RIS-BS channel, one effective channel can be represented by the product of a vector, termed the small-timescale effective channel, and the other effective channel. Once the first effective channel is estimated, subsequent effective channels can be recovered by estimating only lower-dimensional small-timescale effective channels. For large-timescale channel estimation, we utilize the state-of-the-art (SOTA) method, named PW-CLRA \cite{Lee2025}, which performs well in near-field channels. 
     
    \item Our primary contributions are twofold: i) We introduce a piecewise beam training method that effectively manipulates the RIS, and ii) We propose an efficient channel estimation method based on the pilot signals processed through this piecewise beam training, which is formulated as a multiple least squares (Multi-LS) problem. Additionally, we present a theoretical analysis of our RIS design, which offers guidelines for hyperparameter selection. The proposed channel estimation method, based on the time-scaling property, is designated as {\bf 2TCE-TSP}.

    \item Simulations confirm the superiority of the 2TCE-TSP method across various channels, achieving a $38\%$ to $78\%$ reduction in pilot overhead compared to the SOTA method. And, our approach demonstrates significant performance improvements across all pilot overhead regimes while reducing computational complexity by approximately $98\%$ with $512$ RIS elements compared with the benchmark method.
    
\end{itemize}

\subsection{Organization}
The remainder of this paper is organized as follows. Section II defines the system model for RIS-aided near-field communication systems, including the hybrid beamforming architecture. In Section III, we propose a novel channel estimation method, designated as 2TCE-TSP, which consists of two phases: i) large-timescale channel estimation and ii) small-timescale channel estimation. Section IV provides an analysis of the proposed 2TCE-TSP method regarding accuracy, computational complexity, and pilot overhead. Section V presents simulation results, and Section VI concludes the paper.



{\em Notations.} Let $[N_1: N_2]\eqdef \{N_1,N_1+1,...,N_2\}$ for any integer $N_1, N_2$ with $N_2>N_1$. For $N_1=1$, this notation simplifies to $[N_2]$. We use $\xv$ and $\Am$ to denote a column vector and matrix, respectively. Also, $\circ$ denotes the Hadamard product. Given a $M \times N$ matrix $\Am$, let $\Am(i,:)$ and $\Am(:,j)$ denote the $i$-th row and $j$-th column of $\Am$, respectively. Given $m < M$ and $n < N$, let $\Am([m],:)$ and $\Am(:,[n])$ denote the submatrices by taking the first $m$ rows and $n$ columns of $\Am$, respectively. Also, $\mbox{Rank}\left(\Am\right)$, $\Am^{\herm}$,$\|\Am\|_2$, and $\|\Am\|_F$ denote the rank, the Hermitian transpose, the $\ell_2$-norm, and the Frobenius norm of $\Am$, respectively. Given a vector $\vv$, ${\vv}^{*}$ denotes the complex conjugate vector of $\vv$ and ${\rm diag}(\vv)$ denotes a diagonal matrix whose $\ell$-th diagonal element corresponds to the $\ell$-th element of $\vv$. Without loss of generality, it is assumed that in the diagonal matrix resulting from the eigenvalue decomposition, the diagonal elements, which correspond to the eigenvalues, are arranged in descending order based on their absolute values.

\begin{figure}[t]
\centering
\includegraphics[width=1.0\linewidth]{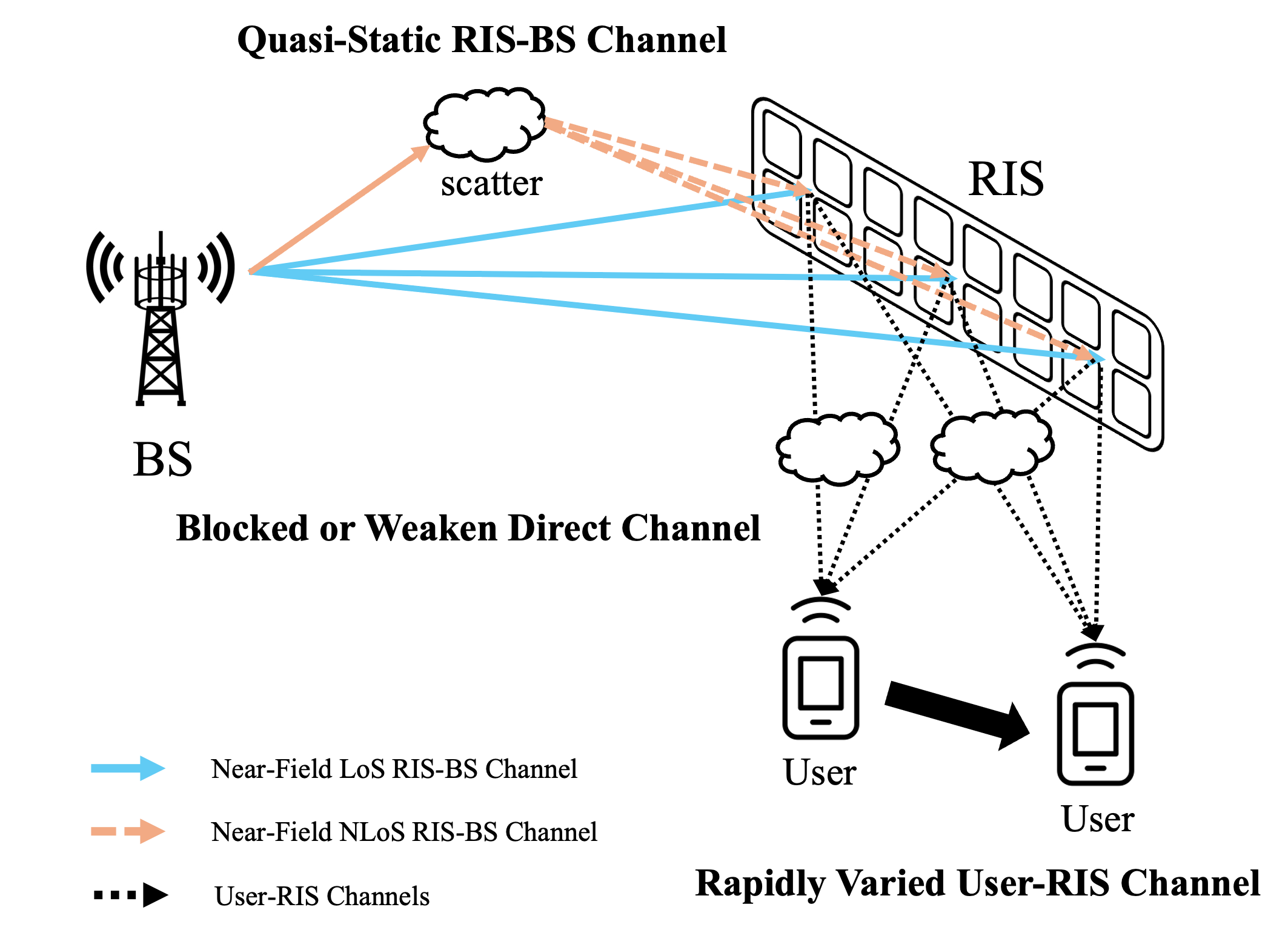}
\caption{The description of the RIS-aided near-field communication systems.}
\end{figure}

\section{System Model}\label{sec:System_Model}%
As illustrated in Fig. 1, we investigate a millimeter-wave (mmWave) multiple-input multiple-output (MIMO) system operating in time division duplex (TDD) mode, wherein a base station (BS) serves mobile users. The BS is equipped with $N$ antennas, while each user is equipped with a single antenna. Given the substantial path loss in mmWave or THz communications, the direct channel between the BS and the users is frequently obstructed or weakened. To address this challenge, we posit that a reconfigurable intelligent surface (RIS) is strategically placed between the BS and the users. This RIS facilitates communication by adaptively manipulating the incident waves. It consists of $M$ antennas connected to an $M$-port single connected reconfigurable impedance network, commonly known as a diagonal RIS. 

In this paper, we aim to develop a channel estimation method for the implementation of  RIS-aided near-field communications. In TDD systems, the downlink channels are derived from the uplink channel estimation due to channel reciprocity. Consequently, our channel estimation method will be articulated based on the uplink channel estimation protocol. To enhance clarity, we will elucidate the proposed channel estimation method within the context of a single-user setting. However, this method can be readily extended to a multi-user scenario through the use of orthogonal pilots.

\subsection{Signal Model}
To reduce complexity and power consumption, it is assumed that the BS is equipped with a hybrid beamforming architecture. In this configuration, all receive antennas share a limited number of radio frequency (RF) chains, represented as $N_{\rm RF}\ll N$. Consequently, the uplink signal from the user to the BS via the RIS can be expressed in its baseband representation:
\begin{align}
    \yv &= \Wm^{\rm RF}\left(\Hm^{\rm RB}\mbox{diag}(\vv)\hv^{\rm UR}s + \nv\right)\nonumber\\
    &\stackrel{(a)}{=} \Wm^{\rm RF}\left(\Hm^{\rm eff}\vv{s} + \nv\right)\in\CC^{N_{\rm RF}\times 1},\label{eq:signalmodel}
\end{align} where $s\in\CC$ represents the transmit signal at the user, constrained by the transmit power as $|s|^2=P$, $\hv^{\rm UR}\in\CC^{M\times 1}$ denotes the channel from the user to the RIS, referred to as the User-RIS channel, $\vv\in\CC^{M\times 1}$ represents the RIS reflection vector, $\Hm^{\rm RB}\in\CC^{N\times M}$ denotes the channel from the RIS to the BS, termed the RIS-BS channel, and $\Wm^{\rm RF}\in\CC^{N_{\rm RF}\times N}$ signifies the analog combiner at the BS. Furthermore, the equality in (a) is derived from the following relationship:
\begin{equation}
    \Hm^{\rm RB}\mbox{diag}(\vv)\hv^{\rm UR} = \Hm^{\rm RB}\mbox{diag}(\hv^{\rm UR})\vv = \Hm^{\rm eff}\vv,\label{eq:effetive}
\end{equation} where $\Hm^{\rm eff}\eqdef \Hm^{\rm RB}\mbox{diag}(\hv^{\rm UR})\in\CC^{N\times M}$ is referred as the {\em effective channel} in this paper. It is noteworthy that both the analog combiner $\Wm^{\rm RF}$ and the RIS reflection vector $\vv$ comply with the constant modulus constraint, where each element possesses the same magnitude as specified below:
\begin{equation}
    |\Wm^{\rm RF}(i,j)| = 1/\sqrt{N}\mbox{ and }|\vv(m)| = 1, \label{eq:CM}
\end{equation} for all $i\in[N^{\rm RF}]$, $j\in[N]$, and $m\in[M]$.


\subsection{Channel Model}\label{subsec:CM}
In relevant studies \cite{Yang2024,Yu2023}, the RIS-BS channel, denoted as $\Hm^{\rm RB}$, has been modeled as a rank-1 channel matrix, based on the dominant line-of-sight (LoS) path and the far-field approximation in mmWave systems. However, as the number of BS and RIS antennas increases, it is essential to adopt the near-field channel for more accurate modeling. Under this assumption, as noted in \cite{Mu2024,Lv2024,Lee2025}, the RIS-BS channel exhibits a {\em high-rank} characteristic when the distance between them is below a certain threshold, significantly enhancing the DoF for multi-stream or multi-user communications. However, this high-rank nature leads to heavy pilot overhead for channel estimation. Motivated by this, we aim to develop an efficient channel estimation method for RIS-aided near-field systems characterized by high-rank effective channels.

\subsubsection{RIS-BS Channel}
Beyond the conventional far-field approximation, which relies on uniform planar wavefronts (UPW), a more realistic near-field assumption must be considered for modeling the RIS-BS channel. This near-field model is based on non-uniform spherical wavefronts (NUSW) \cite{Lu2024} and is particularly relevant when the physical distance between the BS and the RIS is shorter than a critical threshold. When the number of antennas at both the BS and RIS becomes extremely large, this threshold is defined as the MIMO advanced Rayleigh distance (MIMO-ARD) \cite{Lu2023}.  Given a realistic double-sided LoS channel matrix, denoted as $\Am\in\CC^{N\times M}$, the MIMO-ARD is computed as 
\begin{equation}
    {\ell}^{\rm ARD} = \frac{4D^{\rm BS}D^{\rm RIS}}{\lambda_{\rm c}} ,\label{eq:ARD}
\end{equation} where $D^{\rm BS}$ and $D^{\rm RIS}$ denote the antenna apertures of the BS and the RIS, respectively, and $\lambda_{\rm c}$ represents the system wavelength. For instance,  when the number of antennas of the BS and the RIS are respectively given by $N=128$ and $M=512$ comprising the uniform linear array (ULA) with $f_{\rm c} = 100$ GHz, the MIMO-ARD is approximately $200{\rm m}$. Therefore, if the physical distance between the BS and the RIS is shorter than $200{\rm m}$, the RIS-BS channel exhibits near-field characteristics. In this scenario, the $(n,m)$-th entry of $\Am$ is defined as:
\begin{equation}
    \Am(n,m) = \frac{1}{r_{n,m}^{\rm RB}}e^{j{k_{\rm c}}r_{n,m}^{\rm RB}},\; n \in [N], m \in [M],
\end{equation} where $k_{\rm c} = 2\pi{f_{\rm c}}/c$ is the wave number corresponding to the center carrier frequency with the speed of the light $c$, and $r_{n,m}^{\rm RB}$ denotes the physical distance between the $n$-th BS antenna component and the $m$-th RIS antenna component. Note that the free-space path loss is normalized as $\frac{1}{r_{n,m}^{\rm RB}}$ in this paper.

Given the extremely large antenna aperture of the RIS, it is crucial to consider the effects of nearby blockages that may impede signal incidence on specific regions of the RIS \cite{Yu2023}. To characterize this effect, we define a visual region (VR) matrix, denoted as $\Fm\in\{0,1\}^{N\times M}$. If the path from the $m$-th RIS antenna to the $n$-th BS antenna is obstructed, the $(n,m)$-th entry of the VR matrix is set to zero; otherwise, it is set to one. Given that blockages are randomly distributed in practice, we model each element in the VR matrix as an independently and identically distributed (i.i.d.) random variable, taking the value of one with probability $p$. By modeling the effective channel under i.i.d environments, we are able to assess the average performance of the proposed channel estimation method applicable to general near-field channels. Consequently, the RIS-BS channel, considering the near-field assumption and the VR, can be expressed as follows:
\begin{equation}\label{eq:RB}
    \Hm^{\rm RB} = \Am\circ\Fm + \Hm_{\rm nlos}^{\rm RB} \in\CC^{N\times M},
\end{equation} where $\circ$ denotes the Hadamard product, and $\Hm_{\rm nlos}^{\rm RB}\in\CC^{N\times M}$ represents the non line-of-sight (NLoS) RIS-BS channel comprising $L^{\rm RB}$ signal paths.

\subsubsection{User-RIS Channel}
We also incorporate the near-field assumption in modeling the User-RIS channel, denoted as $\hv^{\rm UR} \in \CC^{M\times 1}$. Notably, if the physical distance between the RIS and the user is less than
\begin{equation}
    {\ell}^{\rm RD} = \frac{2(D^{\rm RIS})^2}{\lambda_{\rm c}},\label{eq:RD}
\end{equation} referred to as the MIMO Rayleigh distance (MIMO-RD) \cite{Lu2023}, the User-RIS channel exhibits near-field characteristics. Let $\av\in\CC^{M\times 1}$ represent the single-sided LoS channel vector. In the near-field region, the $m$-th element of the LoS channel can be modeled as follows:
\begin{equation}
    \av(m) = \frac{1}{r_{m}^{\rm UR}}e^{j{k_{\rm c}}r_{m}^{\rm UR}},\; m \in [M],
\end{equation} where $r_{m}^{\rm UR}$ represents the physical distance between the $m$-th RIS antenna and the user. Furthermore, the VR vector for the User-RIS channel is defined as $\fv\in\CC^{M\times 1}$. Likewise, the User-RIS channel, considering the near-field assumption and the VR, is ultimately represented as follows:
\begin{equation}
    \hv^{\rm UR} = \av\circ\fv + \hv_{\rm nlos}^{\rm UR},\label{eq:UR}
\end{equation} where $\hv_{\rm nlos}^{\rm UR}\in\CC^{M\times 1}$ denotes the NLoS channel for the User-RIS channel, comprising $L^{\rm UR}$ signal paths.

\vspace{0.1cm}
\begin{remark}
    It is important to note that in mmWave communications, the dominant signal path is the LoS path, where significant reflection losses result in NLoS paths experiencing average attenuation that exceeds $10$ dB compared to the LoS path \cite{Priebe2013}. Furthermore, the number of NLoS paths is typically small due to the severe path loss. Consequently, the characteristics of the effective channel $\Hm^{\rm eff}$ are primarily derived from the dominant LoS channels, specifically $\Am\circ\Fm$ and $\av\circ\fv$. Nonetheless, for practical applications, the effects of NLoS channel must also be considered. According to the near-field NLoS channel model that incorporates the VR \cite{chen2024, Wu2025}, we appropriately design the near-field NLoS channels, namely, $\Hm_{\rm nlos}^{\rm RB}$ and $\hv_{\rm nlos}^{\rm UR}$ in \eqref{eq:RB} and \eqref{eq:UR}, respectively.
\end{remark}


\subsection{Rank Analysis of Channel Model}\label{subsec:Rankanalysis}
\begin{figure}[t]
\centering
\includegraphics[width=1.0\linewidth]{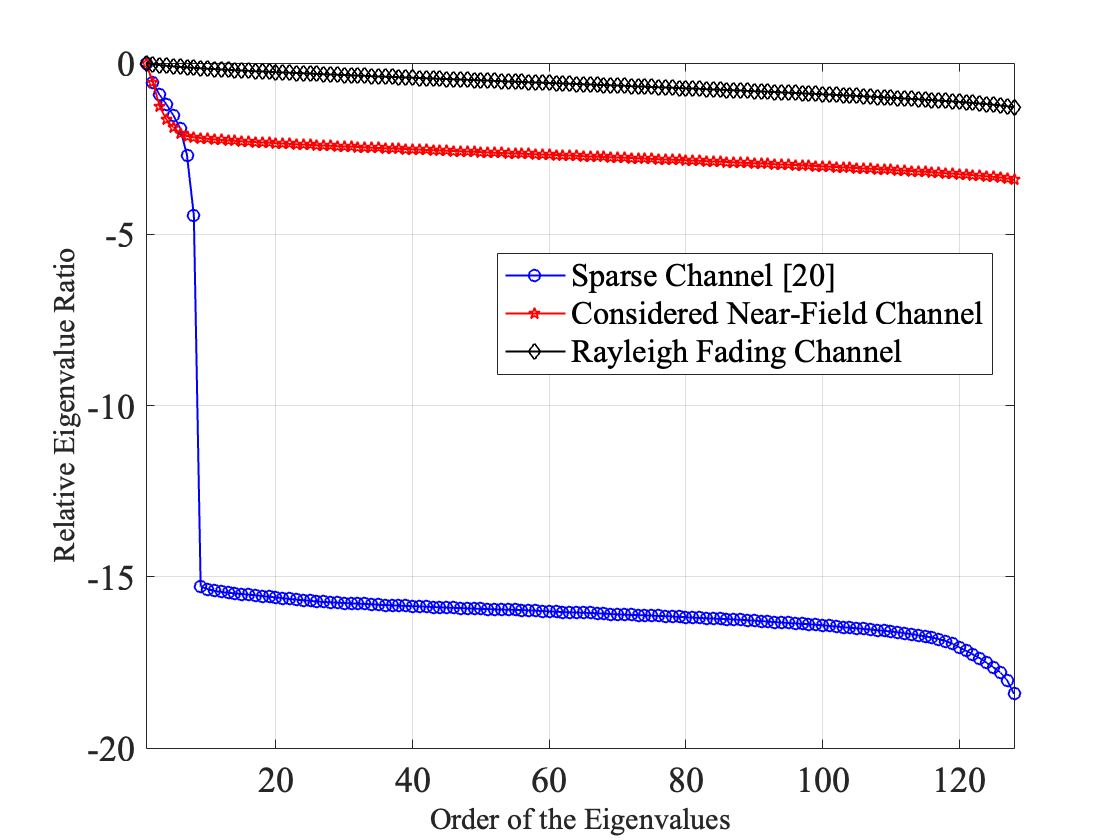}
\caption{The relative eigenvalue ratio of the effective channels as a function of orders across various channel models. $N=128$, and $M=512$.}
\end{figure}

We analyze the effective channel matrix $\Hm^{\rm eff}$ by leveraging the relative eigenvalue ratio defined as:
\begin{equation}
\zeta_n\left(\Hm^{\rm eff}\right)\eqdef\log_{10}\left(|\lambda_n|/|\lambda_1|\right),
\end{equation} where $\lambda_n$ denotes the $n$-th largest eigenvalue of $\Hm^{\rm eff}\left(\Hm^{\rm eff}\right)^{\herm}$, for $n\in[N]$. Thus, $\zeta_n\left(\Hm^{\rm eff}\right)$ represents the ratio between the $n$-th largest eigenvalue and the largest eigenvalue of $\Hm^{\rm eff}\left(\Hm^{\rm eff}\right)^{\herm}$. This enables us to analyze the rank of the effective channel. To illustrate the characteristics of the effective channel, we conduct experiments measuring the relative eigenvalue ratio across various channel models, with simulation environments consistent with those considered in our simulations. Fig. 2 depicts the relative eigenvalue ratio as a function of the order of the eigenvalues. Notably, as the order increases, the corresponding eigenvalues diminish. In the sparse channel model \cite{Yang2024}, the ratio exhibits a significant phase transition at a very low order, implying that most eigenvalues in the sparse channel approach zero compared to the largest eigenvalue. Consequently, the channel behaves as a low-rank matrix, which can be efficiently estimated using existing methods \cite{chen2023channel,schroeder2022channel,Yang2024,Lee2024near}. In contrast, the ratio for the near-field channel model shows a lower value but similar behavior to the Rayleigh fading channel model, which benefits from rich multi-path assumptions. This suggests that the effective channel under consideration exhibits a high-rank matrix.

\begin{table*}[ht]
\caption{Comparisons of 2TCE-PWCLRA and 2TCE-TSP}
\setlength{\tabcolsep}{5pt}
\renewcommand{\arraystretch}{1.5}

 \centering

 \begin{tabular}
 {|c||c|c|c|c|c|c|c}
\toprule[1.5pt]
\hline
   & & $t=0$ & $t=1$ & $t=2$ & $\cdots$ & $t=T-1$\\ 
\hline
\multirow{2}{*}{2TCE-PWCLRA} & Estimated Channel & $\{\hat{\Sm}_q, \Tm_{[q,0]}: q \in [Q]\}$ & 
$\{\Tm_{[q,1]}: q\in[Q]\}$ & $\{\Tm_{[q,2]}: q\in[Q]\}$ & $\cdots$ & $\{\Tm_{[q,T-1]}: q\in[Q]\}$  \\ 
\cline{2-7}
&Pilot Overhead & $Q(N/N_{\rm RF})+M$ & $M$ & $M$ & $\cdots$ & $M$  \\ 
\hline
 \multirow{2}{*}{2TCE-TSP} &Estimated Channel & $\{\Hm_{[q,0]}^{\rm pw}: q \in [Q]\}$ &$\{\dv_{[q,1]}: q\in[Q]\}$& $\{\dv_{[q,2]}: q\in[Q]\}$ & $\cdots$ & $\{\dv_{[q,T-1]}: q\in[Q]\}$ \\
 \cline{2-7}
 &Pilot Overhead & $Q(N/N_{\rm RF})+M$ & $M/N_{\rm RF}$ & $M/N_{\rm RF}$ & $\cdots$ &  $M/N_{\rm RF}$ \\ 
\hline
\bottomrule[1.5pt]
\end{tabular}
\end{table*}

\section{Proposed Channel Estimation Method}\label{sec:proposed}
In this section, we introduce an efficient channel estimation method tailored for a near-field RIS-assisted MU-MIMO system. The proposed method exploits the characteristics of the channel model described in Section~\ref{subsec:CM}. Importantly, the BS and the RIS are positioned at fixed locations, while the mobile users move rapidly, leading to an {\em asymmetry} between the channel coherence times of the RIS-BS channel and the User-RIS channel. The RIS-BS channel is characterized as quasi-static, necessitating estimation over a larger timescale. Conversely, the User-RIS channel exhibits time variation due to user mobility, requiring estimation on a smaller timescale. Throughout this paper, the coherence times for the RIS-BS and User-RIS channels are denoted as $T^{\rm RB}$ and $T^{\rm UR}$, respectively. To facilitate exposition, we assume that the ratio $T=T^{\rm RB}/T^{\rm UR}$ is a positive integer. Within the longer coherence time $T^{\rm RB}$, there exists a sequence of $T$ User-RIS channels, denoted as $\hv_{t}^{\rm UR}$ for $t\in[0:T-1]$. Consequently, from  \eqref{eq:effetive}, the sequence of effective channels is represented as 
\begin{equation}
    \Hm_{t}^{\rm eff}=\Hm^{\rm RB}\mbox{diag}(\hv_t^{\rm UR}),\; t\in[0:T-1].
\end{equation}

From this, we derive the {\em time-scaling property}, which indicates that  each $\Hm_t^{\rm eff}$ can be expressed as the product of the initial effective channel $\Hm_0^{\rm eff}$ and a time-specific $M\times 1$ vector:
\begin{align}
    \Hm_t^{\rm eff} &= \Hm^{\rm RB}\mbox{diag}(\hv_0^{\rm UR})\mbox{diag}(\dv_t)\nonumber\\
    & = \Hm_0^{\rm eff}\mbox{diag}(\dv_t),\label{eq:timescaling}
\end{align} where the small-timescale effective channel is defined as:
\begin{equation}
    \dv_t \eqdef \mbox{diag}(\hv_0^{\rm UR})^{-1}\hv_t^{\rm UR}\in\CC^{M\times 1}.\label{eq:smalltimescale}
\end{equation} By leveraging this property, we present a two-timescale channel estimation framework, referred to as {\bf 2TCE}, which consists of two phases:
\begin{itemize}
    \item Large-timescale channel estimation: By employing the SOTA channel estimation method as outlined in \cite{Lee2025}, we estimate the large-timescale effective channel matrix during the initial time block, which is denoted as $\hat{\Hm}_0^{\rm eff}$. 
    \item Small-timescale channel estimation: As the primary contribution of this paper, we estimate the subsequent small-timescale effective channels by utilizing the initially estimated large-timescale effective channel. The estimated channels are denoted as  $\{\hat{\dv}_t: t\in[T-1]\}$. 
\end{itemize} As outlined in \cite{Hu2021}, we note that the pilot overhead required for estimating the small-timescale effective channels, consisting of $M$ channel parameters, is significantly smaller than that necessary for the large-timescale effective channel, which encompasses $NM$ channel parameters. Notably, the reduction in overhead becomes more pronounced as the number of antennas at the BS increases. Furthermore, the strategy outlined in 2TCE can be directly applied to the two-timescale beamforming optimization and RIS design, arises that have been explored in the current literature \cite{Palmucci2023,Zhi2023}.

\subsection{Large-Timescale Channel Estimation}
We elucidate the piecewise low-rank approximation, initially introduced in \cite{Lee2025}, as the primary technique for large-timescale channel estimation. In the context of RIS-assisted near-field communications, both the RIS-BS channel and the User-RIS channel typically exhibit near-field characteristics due to the significantly enlarged antenna aperture of the RIS, which facilitates high data rates 
\cite{Mu2024,Lv2024,Zhou2015}. This near-field property leads to a loss of channel sparsity or low-rankness, which is essential for effectively reducing pilot overhead in existing methods \cite{chen2023channel,schroeder2022channel,Yang2023,Yang2024,Lee2024near}. In \cite{Lee2025}, to address the high-rank characteristic of the RIS-BS channel $\Hm^{\rm RB}$, the effective channel in the initial time block is divided into $Q$ distinct {\em piecewise effective channels} $\{\Hm_{[q,0]}^{\rm pw}:q\in[Q]\}$, with the condition that $M_{\rm sub}Q =M$, as follows:
\begin{equation}
    \Hm^{\rm eff}_{0} = \begin{bmatrix}
        \Hm_{[1,0]}^{\rm pw} &\cdots& \Hm_{[Q,0]}^{\rm pw}
    \end{bmatrix}.\label{eq:pweff}
\end{equation} Herein, each piecewise effective channel is defined as
\begin{equation}
    \Hm_{[q,0]}^{\rm pw} = \Hm_{q}^{\rm RB}\mbox{diag}(\hv_{[q,0]}^{\rm UR})\in\CC^{N\times M_{\rm sub}},
\end{equation} where
\begin{equation}
    \Hm_{q}^{\rm RB} = \Hm^{\rm RB}(:,\Mc_q)\mbox{ and }\hv_{[q,0]}^{\rm UR} = \hv_{0}^{\rm UR}(\Mc_q),
\end{equation} and $\Mc_q = \{1+(q-1)M_{\rm sub}:qM_{\rm sub}\}$. It was demonstrated in \cite{Lee2025} that each piecewise effective channel can be accurately approximated as the product of low-rank matrices. Leveraging this observation, each piecewise effective channel can be expressed as follows:
\begin{equation}
    \Hm_{[q,0]}^{\rm pw} = \Sm_q\Tm_{[q,0]},\label{eq:pwdec}
\end{equation} where $\Sm_q$ is an $N\times r_q$ matrix whose $r_q$ columns span the column space of $\Hm_q^{\rm RB}$, and $\Tm_{[q,0]}\in\CC^{r_q\times M_{\rm sub}}$ is the coefficient matrix associated with $\Sm_q$.

The SOTA channel estimation method, referred to as PW-CLRA and proposed in \cite{Lee2025}, can effectively estimate the piecewise effective channels. By leveraging this method, the effective channel in the initial time block can be estimated, denoted as follows:
\begin{align}
    \hat{\Hm}_0^{\rm eff} &= \begin{bmatrix}
        \hat{\Hm}_{[1,0]}^{\rm pw} &\cdots& \hat{\Hm}_{[Q,0]}^{\rm pw}
    \end{bmatrix}.\label{eq:initial}
\end{align} Given that $\Sm_{q}$ is accurately approximated as a low-rank matrix, the PW-CLRA method estimates the piecewise effective channels while ensuring a short pilot overhead in the following manner:
\begin{align}
    \hat{\Hm}_{[q,0]}^{\rm pw} &= \hat{\Sm}_q\hat{\Tm}_{[q,0]}={\Hm}_{q}^{\rm pw} + \Deltam_{[q,0]},
    \label{eq:estimatedeffect}
\end{align} where $\hat{\Sm}_q\in\CC^{N\times \hat{r}_q}$ and $\hat{\Tm}_{[q,0]}\in\CC^{\hat{r}_q\times M_{\rm sub}}$ are estimated using piecewise low-rank approximation with the estimated rank $\hat{r}_q$ \cite{Lee2025}, and $\Deltam_{[q,0]}\in\CC^{N\times M_{\rm sub}}$ denotes the channel estimation error.

In future wireless communication systems, a dense mobile user scenario (i.e., $K\gg1$) is typically assumed. Consequently, the PW-CLRA algorithm is capable of mitigating the channel estimation error by leveraging the multi-user gains to estimate the common subspace $\Sm_q$. For the sake of brevity in describing the small-timescale channel estimation in Section~\ref{subsec:STCE}, we will disregard the impact of channel estimation errors, specifically as $\hat{r}_q = r_q$ and ${\Deltam}_{[q,0]} = \mbox{0}$. However, these effects will be thoroughly investigated in our simulations.  As derived in \cite{Lee2025}, the minimum pilot overhead of the PW-CLRA is expressed as 
\begin{equation}
    Q(N/N_{\rm RF}) + M,\label{eq:overhead_PW}
\end{equation} where the first term $Q(N/N_{\rm RF})$ and the second term $M$ are required for the estimations of $\{\Sm_q:q\in[Q]\}$ and $\{\Tm_{[q,0]}:q\in[Q]\}$, respectively. By selecting the hyperparameter $Q$, one can effectively manage the tradeoff between estimation accuracy and pilot overhead.

\subsection{Small-Timescale Channel Estimation}\label{subsec:STCE}
Based on the estimated common subspace, denoted as $\hat{\Sm}_{q}$ in \eqref{eq:estimatedeffect}, we can model the subsequent effective channels as follows:
\begin{equation}
    \Hm_{[q,t]}^{\rm pw} = \hat{\Sm}_{q} \Tm_{[q,t]},\; t \in [T-1].
\end{equation}  During the time block 
$t\geq 1$, we can recover the effective channel by estimating only the 
coefficient matrix $\Tm_{[q,t]}$ for $q \in [Q]$. This can be efficiently derived using the least square (LS) or joint optimization (JO) solution, outlined in \cite{Lee2025}. This two-phase channel estimation method based on PW-CLRA is referred to as {\bf 2TCE-PWCLRA}. In this method, compared to applying the PW-CLRA to each time block, the minimum pilot overhead per time block can be reduced to $M$. Nonetheless, the reduced pilot overhead may become unaffordable as the number of RIS elements increases considerably.

Motivated by the aforementioned observations, we propose an efficient small-timescale channel estimation method that achieves a pilot overhead below $M$, which constitutes the primary contribution of this paper. By leveraging the time-scaling property as defined in \eqref{eq:timescaling}, we derive the time-scaling property in a piecewise manner as follows:
\begin{align}
    \Hm_{[q,t]}^{\rm pw} &= \Hm_q^{\rm RB}\mbox{diag}(\hv_{[q,t]}^{\rm UR})= \Hm_{[q,0]}^{\rm pw}\mbox{diag}(\dv_{[q,t]}),\label{eq:pwtimescaling}
\end{align} where the piecewise small-timescale effective channel is defined as:
\begin{equation}
    \dv_{[q,t]}\eqdef\mbox{diag}(\hv_{[q,0]}^{\rm UR})^{-1}\hv_{[q,t]}^{\rm UR}\in\CC^{M_{\rm sub}\times 1}.
\end{equation} Notably, $\Hm_{[q,0]}^{\rm pw}$ has been already estimated in the initial time block. In the subsequent time blocks, we will estimate the piecewise small-timescale effective channels $\{\dv_{[q,t]}:q\in[Q],t\in[T-1]\}$. This estimation process is significantly simpler than directly estimating the piecewise effective channels  $\{\Hm_{[q,t]}^{\rm pw}:q\in[Q],t\in[T-1]\}$ using the SOTA method or the coefficient matrices $\{\Tm_{[q,t]}^{\rm pw}:q\in[Q],t\in[T-1]\}$ via 2TCE-PWCLRA.

To facilitate this, we first introduce a novel beam training method designed to derive suitable measurements for our channel estimation approach in Section~\ref{subsubsec:PBT}. Subsequently, in Section~\ref{subsubsec:PMT}, we delineate our channel estimation method, referred to as {\bf 2TCE-TSP}.

\subsubsection{Piecewise Beam Training}\label{subsubsec:PBT}
For each time block $t$, the proposed beam training proceeds with $B\leq M_{\rm sub}$ subframes, each containing $Q$ pilot symbols. According to the signal model defined in \eqref{eq:signalmodel}, the BS receives the following signal during the $i$-th pilot transmission within the subframe $b$:
\begin{align}
    \yv_{[b,i,t]} &= \Wm^{\rm RF}\left(\Hm^{\rm RB}\mbox{diag}(\nuv_{[b,i]})\hv_t^{\rm UR}s + \nv_{[b,i,t]}\right),\label{eq:uplinkpilot}
\end{align}  where $\nuv_{[b,i]}$ and $\nv_{[b,i,t]}$ denote the RIS reflection vector and the additive noise, respectively. By dividing by the known pilot symbol, we can obtain:
\begin{align}
    \tilde{\yv}_{[b,i,t]}& = \Wm^{\rm RF}\Hm_t^{\rm eff}\nuv_{[b,i]} + \tilde{\nv}_{[b,i,t]}\nonumber\\
    &\stackrel{(a)}{=} \Wm^{\rm RF}\sum_{q=1}^{Q}\Hm_{[q,t]}^{\rm pw}\nuv_{[b,i]}(\Mc_q) + \tilde{\nv}_{[b,i,t]},\label{eq:ulpilot}
\end{align} where $\tilde{\nv}_{[b,i,t]}=\frac{1}{s}\Wm^{\rm RF}\nv_{[b,i,t]}$ and (a) follows from the definition of the piecewise effective channels in \eqref{eq:pweff}.

\begin{figure}[t]
\centering
\includegraphics[width=0.8\linewidth]{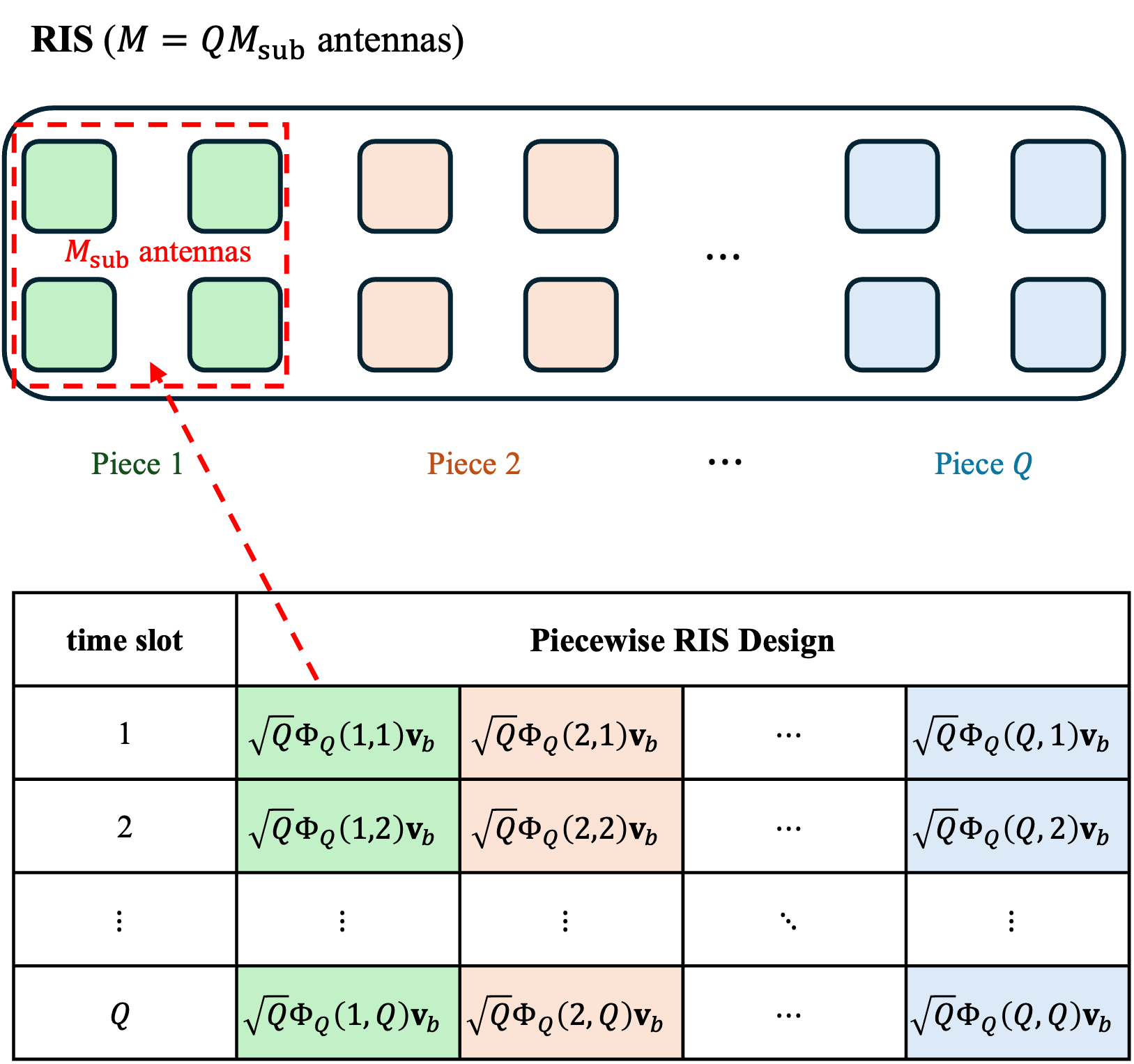}
\caption{The description of the proposed piecewise beam training at the $b$-th subframe, comprising $Q$ pilot symbols, where $\vv_{b}$ is the $M_{\rm sub}$-dimensional vector that depends on the subframe.} 
\end{figure}
To derive piecewise observations via beam training, we construct the RIS reflection vector in a piecewise manner as illustrated in Fig. 3, adhering to the constant modulus constraint as defined in \eqref{eq:CM}:
\begin{equation}
    \nuv_{[b,i]}(\Mc_q) = \sqrt{Q}\Phim_{Q}(q,i)\vv_b\in\CC^{M_{\rm sub}\times 1}
\end{equation} for $q \in [Q]$, where $\Phim_Q$ is a $Q\times Q$ unitary matrix and $\vv_b$ is a $M_{\rm sub}\times 1$ vector. For each $b \in [B]$, we design $\vv_b$ by sequentially selecting the $b$-th column of an $M_{\rm sub}\times M_{\rm sub}$ unitary matrix, such as discrete Fourier transform (DFT) matrix or Hadamard matrix. In Section~\ref{subsec:theoanalysis}, we will conduct a theoretical analysis to validate the effectiveness of our RIS design. By concatenating the $Q$ received signals $\{\tilde{\yv}_{[b,i,t]}:i\in[Q]\}$, we can define:
\begin{align}
    \Zm_{[b,t]} &= \frac{1}{\sqrt{Q}}\begin{bmatrix}
        \tilde{\yv}_{[b,1,t]} &\cdots& \tilde{\yv}_{[b,Q,t]}
    \end{bmatrix}\Phim_Q^{\herm}\nonumber\\
    &\stackrel{(a)}{=}\Wm^{\rm RF}\begin{bmatrix}
        \Hm_{[1,t]}^{\rm pw}\vv_b &\cdots&\Hm_{[Q,t]}^{\rm pw}\vv_b
    \end{bmatrix} + \tilde{\Um}_{[b,t]},
\end{align} where (a) follows from the fact that $\Phim_Q$ is a unitary matrix and
\begin{equation}
    \tilde{\Um}_{[b,t]}=\frac{1}{\sqrt{Q}}\begin{bmatrix}
      \tilde{\nv}_{[b,1,t]} &\cdots&  \tilde{\nv}_{[b,Q,t]}
    \end{bmatrix}\Phim_Q^{\herm}.\label{eq:beamforming gain}
\end{equation} Ultimately, we obtain the piecewise observations as follows:
\begin{align}
    \zv_{[b,q,t]} &\eqdef \Zm_{[b,t]}(:,q) = \Wm^{\rm RF}\Hm_{[q,t]}^{\rm pw}\vv_b + \tilde{\uv}_{[b,q,t]},\label{eq:pwobserve}
\end{align} where $\tilde{\uv}_{[b,q,t]} = \tilde{\Um}_{[b,t]}(:,q)$.

In the context of the proposed piecewise beam training, deriving an optimal analog combiner  $\Wm^{\rm RB}$ to maximize estimation accuracy presents a challenge. As a practical approach, we design the analog combiner using a fixed unitary matrix (e.g., the DFT matrix or the Hadamard matrix), which satisfies the constant modulus constraint as outlined in \eqref{eq:CM} and does not incur the issue of noise amplification.

\vspace{0.2cm}
\subsubsection{The Proposed 2TCE-TSP}\label{subsubsec:PMT}
We delineate our approach to efficiently estimate the small-timescale effective channels $\{{\dv}_{[q,t]}: q \in [Q], t \in [T-1]\}$ by utilizing the estimated effective channel from the initial time block, namely, $\hat{\Hm}_{0}^{\rm eff}$ as defined in \eqref{eq:initial}, along with the piecewise observations in \eqref{eq:pwobserve}. Our explanation focuses on the $q$-th piecewise effective channel and the $t$-th time block, while The same procedures will subsequently be applied to the other piecewise effective channels.

Based on the observations in \eqref{eq:pwobserve} and the time-scaling property in \eqref{eq:pwtimescaling}, the small-timescale channels can be recovered through the formulation of a linear inverse problem:
\begin{equation}
   \zv_{[b,q,t]} = \Am_{[b,q]}{\dv}_{[q,t]} + \tilde{\uv}_{[b,q,t]} \in \CC^{N_{\rm RF}\times 1},
\end{equation} where the sensing (or measurement) matrix is formed as:
\begin{equation}\label{eq:Ab}
\Am_{[b,q]} \eqdef \Wm^{\rm RF}\hat{\Hm}_{[q,0]}^{\rm pw}\mbox{diag}(\vv_b)\in\CC^{N_{\rm RF}\times M_{\rm sub}}.
\end{equation} We address this problem by pursuing a least-square (LS) solution:
\begin{equation}
{\dv}_{[q,t]}^{\rm LS}=\argmin_{\dv}\;\left\|\zv_{[b,q,t]}-\Am_{[b,q]}\dv\right\|_F^2.
\end{equation} According to the first-order optimality condition, an optimal solution must satisfy the normal equation:
\begin{equation}
    \left(\Am_{[b,q]}^{\herm}\Am_{[b,q]}\right)\dv = \Am_{[b,q]}^{\herm}\zv_{[b,q,t]}.\label{eq:normal}
\end{equation} From the definition of $\Am_{[b,q]}$, the Gram matrix can be specified as follows:
\begin{align}
    \Am_{[b,q]}^{\herm}\Am_{[b,q]}&=\mbox{diag}(\vv_b^{*}) \left(\Wm^{\rm RF}\hat{\Hm}_{[q,0]}^{\rm pw}\right)^{\herm}\Wm^{\rm RF}\hat{\Hm}_{[q,0]}^{\rm pw} \mbox{diag}(\vv_b)\nonumber\\
    &=\left(\mbox{diag}({\vv}_b^{*})\Um\right)\Lambda\left(\mbox{diag}({\vv}_b^{*})\Um\right)^{\herm}\nonumber\\
    &\stackrel{(a)}{=}\sum_{i=1}^{r}\lambda_i\left({\vv}_b^{*}\circ\uv_i\right)\left({\vv}_b^{*}\circ\uv_i\right)^{\herm}\nonumber\\
    &\stackrel{(b)}{=}\left(\vv_b\vv_b^{\herm}\right)\circ\left(\left(\Wm^{\rm RF}\hat{\Hm}_{[q,0]}^{\rm pw}\right)^{\herm}\Wm^{\rm RF}\hat{\Hm}_{[q,0]}^{\rm pw}\right),\label{eq:Gram}
\end{align} where $r \eqdef \mbox{Rank}(\Wm^{\rm RF}\hat{\Hm}_{[q,0]}^{\rm pw})=\min(N_{\rm RF},\hat{r}_q)$, and
\begin{align}
    \left(\Wm^{\rm RF}\hat{\Hm}_{[q,0]}^{\rm pw}\right)^{\herm}\Wm^{\rm RF}\hat{\Hm}_{[q,0]}^{\rm pw}\eqdef \Um\Lambdam\Um^{\herm}=\sum_{i=1}^{r}\lambda_i\uv_i\uv_i^{\herm}
\end{align} from the eigen-decomposition, where $\{\lambda_i: i \in [r]\}$ are positive eigenvalues. (a) follows from the fact:
\begin{equation*}
    \mbox{diag}({\vv}_b^{*})\Um = \begin{bmatrix}
        {\vv}_b^{*}\circ\uv_1 &\cdots& {\vv}_b^{*}\circ\uv_{M_{\rm sub}}
    \end{bmatrix}\in\CC^{M_{\rm sub}\times M_{\rm sub}},
\end{equation*} and (b) is due to the distributive property of the Hadamard product. From \eqref{eq:Gram} and the rank-inequality of the Hadamard product \cite{ando1987singular}, the rank of the Gram matrix is bounded as follows:
\begin{equation}
\mbox{Rank}\left(\Am_{[q,b]}^{\herm}\Am_{[q,b]}\right)\leq\mbox{Rank}\left(\vv_b\vv_b^{\herm}\right) \min(N_{\rm RF},\hat{r}_q).\label{eq:rankineq}
\end{equation}

When $B=1$, the Gram matrix is rank-deficient, as $\mbox{Rank}\left(\vv_b\vv_b^{\herm}\right)\min(N_{\rm RF},\hat{r}_q)= \min(N_{\rm RF},\hat{r}_q) < M$. Consequently, we cannot seek a unique solution from the normal equation in \eqref{eq:normal}. While an arbitrary solution can be obtained using  QR-decomposition or singular value decomposition (SVD) \cite{strang2000linear}, there is no assurance that this solution will closely approximate the desired one. Therefore, to ensure sufficient rank of the Gram matrix for obtaining a desired solution, it is necessary to increase $B$, albeit at the expense of pilot overhead. With the choice of $B > 1$, we formulate a joint optimization problem, referred to as the {\bf multi-LS problem}, from the $B$ piecewise observations  $\{\zv_{[b,q,t]}: b \in [B]\}$:
\begin{equation}
    \min_{\dv}\; \sum_{b=1}^{B}\left\|\zv_{[b,q,t]}-\Am_{[b,q]}\dv\right\|_F^2.\label{eq:jointpro}
\end{equation} 
From the first-order optimality condition, we derive the normal equation for the multi-LS problem as follows:
\begin{align}
    \left(\sum_{b=1}^{B}\Am_{[b,q]}^{\herm}\Am_{[b,q]}\right)\dv &= \sum_{b=1}^{B}\Am_{[b,q]}^{\herm}\zv_{[b,q,t]}. \label{eq:firstorder}
\end{align} By extending the decomposition as described in \eqref{eq:Gram} into the Gram matrix $\Gm_{q}\eqdef \sum_{b=1}^{B}\Am_{[b,q]}^{\herm}\Am_{[b,q]} \in \CC^{M_{\rm sub}\times M_{\rm sub}}$, we obtain:
\begin{align}
    \Gm_q &= \sum_{b=1}^{B}\sum_{i=1}^{r}\lambda_i\left({\vv}_b^{*}\circ\uv_i\right)\left({\vv}_b^{*}\circ\uv_i\right)^{\herm}\nonumber\\
    &=\sum_{b=1}^{B}\left(\vv_b\vv_b^{\herm}\right)\circ\left(\left(\Wm^{\rm RF}\hat{\Hm}_{[q,0]}^{\rm pw}\right)^{\herm}\Wm^{\rm RF}\hat{\Hm}_{[q,0]}^{\rm pw}\right)\nonumber\\
    &\stackrel{(a)}{=}\left(\Vm_B\Vm_B^{\herm}\right)\circ\left(\left(\Wm^{\rm RF}\hat{\Hm}_{[q,0]}^{\rm pw}\right)^{\herm}\Wm^{\rm RF}\hat{\Hm}_{[q,0]}^{\rm pw}\right),\label{eq:Gm}
\end{align} where $\Vm_B\in\CC^{M_{\rm sub}\times B}$ with $B\leq M_{\rm sub}$ is a full-rank matrix and (a) follows from the fact that $\Vm_B\Vm_B^{\herm} = \sum_{b=1}^{B}\vv_b\vv_b^{\herm}$. According to the rank-inequality in \eqref{eq:rankineq}, the rank of the Gram matrix $\Gm_q$ is bounded as follows:
\begin{align}
    \mbox{Rank}\left(\Gm_q\right)&\leq \mbox{Rank}\left(\Vm_{B}\Vm_{B}^{\herm}\right)\min(N_{\rm RF},\hat{r}_q)\nonumber\\
    &=B\min(N_{\rm RF},\hat{r}_q).\label{eq:upperbound}
\end{align} To satisfy the necessary condition for obtaining a unique solution of the multi-LS problem in \eqref{eq:jointpro}, $B$ should be selected such that
\begin{equation}
    B\geq \left\lceil\frac{M_{\rm sub}}{\min(N_{\rm RF},\hat{r}_q)}\right\rceil=\left\lceil \frac{M}{Q\min(N_{\rm RF},\hat{r}_q)}\right\rceil.\label{eq:necessary}
\end{equation}  Consequently, to satisfy the above necessary condition for all pieces, the minimum number of subframes, denoted as $B_{\rm min}$, should be determined as:
\begin{equation}
    B_{\rm min} =  \max_{q\in[Q]}\;\left\lceil\frac{M}{Q\min(N_{\rm RF},\hat{r}_q)}\right\rceil.\label{eq:PO}
\end{equation} We theoretically proved that the Gram matrix $\Gm_{q}$ is full-rank when $B$ is slightly greater than $B_{\rm min}$ (refer to Theorem 1 in Section~\ref{subsec:theoanalysis}). Consequently, the pilot overhead of the proposed 2TCE-TSP is approximately reduced to $M/N_{\rm RF}$, compared to $M$ for 2TCE-PWCLRA.

Suppose that $\mbox{Rank}(\Gm_{q})=M_{\rm sub}$ with a pilot overhead of approximately $QB_{\rm min}$. Then, we can derive the unique solution of the normal equation as follows:
\begin{equation}
    {\dv}_{[q,t]}^{\rm MLS} = \Gm_q^{-1}\left(\sum_{b=1}^{B}\Am_{[b,q]}^{\herm}\zv_{[b,q,t]}\right). \label{eq:LSsol}
\end{equation} In the intermediate and low SNR regimes, however, the noise amplification resulting from the multiplication by $\Gm^{-1}_{q}$ should not be overlooked. In these regimes, the full rank of $\Gm_{q}$ alone does not guarantee accurate estimation of the multiple-LS solution in \eqref{eq:LSsol}. This is due to the inherent sensitivity of the estimation process to perturbations when the condition number, denoted as $\kappa\left(\Gm_q\right)$, is large. Consequently, a small amount of noise can lead to substantial inaccuracies in the estimates. This argument can be elucidated through the following mathematical derivation:
\begin{align}
    \Delta{\dv}_{[q,t]}^{\rm MLS}&=\Gm_q^{-1}\left(\sum_{b=1}^{B}\Am_{[b,q]}^{\herm}\tilde{\uv}_{[b,q,t]}\right)\nonumber\\
    &\leq \left\|\Gm_q^{-1}\right\|_2\left\|\sum_{b=1}^{B}\Am_{[b,q]}^{\herm}\tilde{\uv}_{[b,q,t]}\right\|_2\nonumber\\
    &=\frac{10^{\kappa\left(\Gm_q\right)}}{\|\Gm_q\|_2}\left\|\sum_{b=1}^{B}\Am_{[b,q]}^{\herm}\tilde{\uv}_{[b,q,t]}\right\|_2, \label{eq:estimationerror}
\end{align} where $\Delta{\dv}_{[q,t]}^{\rm MLS}$ denotes the estimation error of the multiple-LS solution in \eqref{eq:LSsol}.
This demonstrates that the estimation error of the multi-LS solution in \eqref{eq:LSsol} may increase exponentially with the condition number $\kappa\left(\Gm_q\right)$. Thus, one can enhance estimation accuracy by reducing the condition number through the appropriate selection of $B\geq B_{\rm min}$. In our simulations, we choose the $B=2B_{\rm min}$ or $B=3B_{\rm min}$, while ensuring that $QB\ll M$.

\section{Analysis of 2TCE-TSP}

In this section, we analyze the proposed 2TCE-TSP. We begin by providing a theoretical analysis concerning the unique solution of our multi-LS problem. Subsequently, we describe the analysis of the computational complexity and the pilot overhead, particularly about the proposed small-timescale channel estimation method.

\subsection{Theoretical Analysis of Multi-LS Problem}\label{subsec:theoanalysis}
We prove that under a mild condition, $\Gm_{q}$ is full rank when $B$ is selected such that $B\geq B_{\rm min}$. Given the $B$ reflection vectors $\{\vv_{b}: b \in [B]\}$, we define the $B$ subspaces of the vector space $\CC^{M_{\rm sub}}$ as follows:
\begin{equation}
    \Vc({\vv}_{b}^{*})=\mbox{Span}\left(\mbox{diag}({\vv_b^{*}})\uv_1,\mbox{diag}({\vv_b^{*}})\uv_2,...,\mbox{diag}({\vv_b^{*}})\uv_r\right),\nonumber
\end{equation} for $b \in [B]$. Note that for any fixed $b \in [B]$, the vectors $\mbox{diag}({\vv_b^{*}})\uv_i$, $i \in [r]$, are linearly independent. That is, these vectors form the basis of $\Vc({\vv}_b^{*})$, because
\begin{equation}\label{eq:LI}
\mbox{diag}({\vv}_b^{*})(c_1\uv_1+c_2\uv_2+\cdots+c_{r}\uv_r)=\zerov,
\end{equation} only when the coefficients $\{c_i: i\in [r]\}$ are all zeros, due to the full-rankness of the diagonal matrix $\mbox{diag}({\vv}_b^{*})$. To ensure that $\mbox{Rank}(\Gm_{q})=M_{\rm sub}$, 
the collection $\{\Vc({\vv}_{b}^{*}):b \in [B]\}$ must cover the vector space $\CC^{M_{\rm sub}}$, i.e.,
\begin{equation}
    \bigcup_{b=1}^{B} \Vc({\vv}_{b}^{*}) = \CC^{M_{\rm sub}}.\label{eq:Conclusion}
\end{equation}  Below, we provide a theoretical proof demonstrating that this condition is satisfied under a mild assumption, given that $B\geq B_{\rm min}$.

\vspace{0.1cm}
\begin{lemma} For any pair of two distinct subframes $b_1, b_2 \in [B]$ with $b_1\neq b_2$, the $2r$ number of vectors, represented as $\{\mbox{diag}(\vv_{b}^{*})\uv_i:b\in\{b_1,b_2\},i\in[r]\}$, are linearly independent if $\Vc({\vv}_{b_1}^{*})\neq \Vc({\vv}_{b_2}^{*})$.
\end{lemma}
\begin{IEEEproof}
    By construction of $\vv_{b_1}$, $\mbox{diag}(\vv_{b_1}^{*})\Um$ is the unitary matrix. Therefore, $\{\mbox{diag}(\vv_{b_1}^{*})\uv_{m}: m \in [M_{\rm sub}]\}$ are the orthonormal bases of $\CC^{M_{\rm sub}}$. Then, we obtain that for $i \in [r]$,
    \begin{align}
        \mbox{diag}(\vv_{b_2}^{*})\uv_i &= \sum_{j=1}^{M_{\rm sub}} \cv_{i}(j)\mbox{diag}(\vv_{b_1}^{*})\uv_j\\
        &=\mbox{diag}(\vv_{b_1}^{*})\Um \cv_{i},
    \end{align} where $\cv_{i} \in \CC^{M_{\rm sub}\times 1}$ denotes a coefficient vector. Concatenating the coefficient vectors, $\cv_{1},...,\cv_{r}$, we define:
    \begin{align}
        & \Cm = \begin{bmatrix}
                \cv_1 &\cdots& \cv_r
            \end{bmatrix}\nonumber\\
            &=\left(\mbox{diag}(\vv_{b_1}^{*})\Um\right)^{\herm}\begin{bmatrix}
                \mbox{diag}(\vv_{b_2}^{*})\uv_1 &\cdots& \mbox{diag}(\vv_{b_2}^{*})\uv_r
            \end{bmatrix}.
    \end{align} Note that for $k \in [r+1,M_{\rm sub}]$, the $k$-th row of $\Cm$ is represented as follows:
    \begin{equation*}
        \Cm(k,:)=\left(\mbox{diag}(\vv_{b_1}^{*})\uv_{i}\right)^{\herm}\begin{bmatrix}
                \mbox{diag}(\vv_{b_2}^{*})\uv_1 &\cdots& \mbox{diag}(\vv_{b_2}^{*})\uv_r
            \end{bmatrix}.
    \end{equation*} Thus, $\Cm([r+1,M_{\rm sub}],:)$ must not contain an all-zero row if \begin{equation}
        \Vc({\vv}_{b_1}^{*}) \neq \Vc({\vv}_{b_2}^{*}).
    \end{equation} This is because $\Vc(\vv_{b_1}^{*})$ is the orthogonal complement of the subspace, defined as $\mbox{Span}\left(\mbox{diag}(\vv_{b_1}^{*})\uv_{r+1},...,\mbox{diag}(\vv_{b_1}^{*})\uv_{M}\right)$. This analysis implies that if $\Vc({\vv}_{b_1}^{*}) \neq  \Vc({\vv}_{b_2}^{*})$, any vector of the set $\{\mbox{diag}(\vv_{b_2}^{*})\uv_i: i \in [r]\}$ must not be represented as a linear combination of the vectors in $\{\mbox{diag}(\vv_{b_1}^{*})\uv_i: i \in [r]\}$. Also, from \eqref{eq:LI}, $\{\mbox{diag}(\vv_{b_2}^{*})\uv_i: i \in [r]\}$ are linearly independent. This completes the proof.
\end{IEEEproof}

Now we are ready to state our main theorem.
\begin{theorem} Suppose that $B\geq B_{\rm min}$. Then, it is ensured that
\begin{equation}
    \mbox{Rank}(\Gm_{q})=M_{\rm sub},
\end{equation} if $\Vc({\vv}_{b_1}^{*})\neq \Vc({\vv}_{b_2}^{*})$ for any $b_1, b_2 \in [B]$ with $b_1\neq b_2$.
\end{theorem}
\begin{IEEEproof}
    From Lemma 1, we can see that the column space of $\Gm_{q}$ is spanned by the linear combination of $M_{\rm sub}$ linearly independent vectors if $\Vc({\vv}_{b_1}^{*})\neq \Vc({\vv}_{b_2}^{*})$ for any $b_1, b_2 \in [B]$ with $b_1\neq b_2$. This completes the proof.
\end{IEEEproof}

\begin{figure}[t]
\centering
\includegraphics[width=1.0\linewidth]{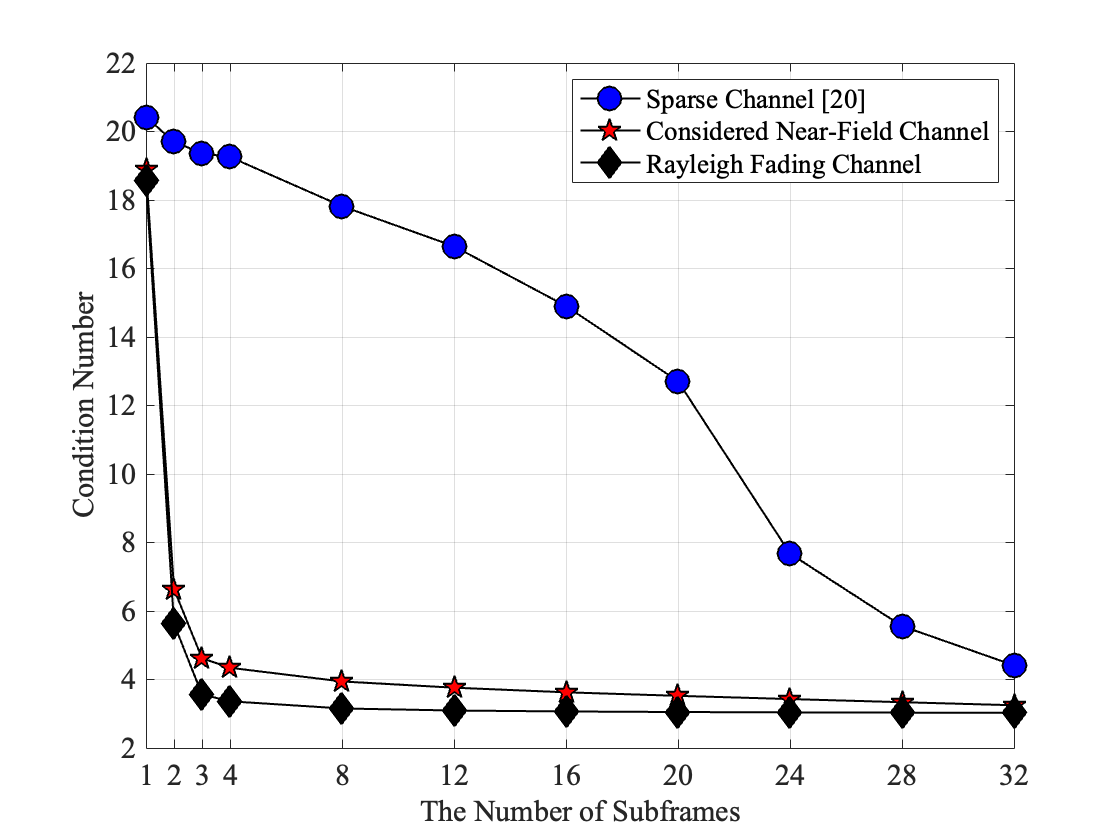}
\caption{The average condition number of the Gram matrices, i.e., $\{\Gm_q:q\in[Q]\}$ as a function of the number of subframes $B$, where $N=128$, $M=512$, $Q=16$, and $N_{\rm RF} = 16$.}
\end{figure}

To validate Theorem 1, we evaluate the condition number of the Gram matrix $\Gm_{q}$ as defined in \eqref{eq:Gm}, which is expressed as:
\begin{equation}
\kappa\left(\Gm_q\right)\eqdef\log_{10}\left(\left\|\Gm_q\right\|_2\left\|\Gm_q^{-1}\right\|_2\right).
\end{equation} The condition number is equivalent to the ratio of the largest eigenvalue to the smallest eigenvalue of $\Gm_q$. When $\kappa\left(\Gm_q\right)$ is large, $\Gm_q$ tends to be a singular matrix, which is referred to as an ill-conditioned matrix. Conversely, $\Gm_q$ becomes a well-conditioned matrix as $\kappa\left(\Gm_q\right)\rightarrow 0$. Fig. 4 illustrates the average condition number of the Gram matrices $\{\Gm_{q}:q\in[Q]\}$, each of which is defined as $\kappa\left(\Gm_q\right)$, as a function of pilot overhead, defined as $QB$. The necessary condition is established as $B\geq B_{\rm min} = 2$, which is derived from \eqref{eq:PO}:
\begin{equation}
    B_{\rm min} = \frac{512}{16\times 16} = 2,\label{eq:hatBmin}
\end{equation} under the assumption that $\hat{r}_q = 16$ for all $q\in[16]$. This assumption implies that the piecewise effective channels in the first time slot represented as $\{\hat{\Hm}_{[q,0]}^{\rm pw}:q\in[16]\}$, are all full-rank channels, consistent with the high-rank effective channel of the considered model. From Fig. 4, we first observe that our theoretical analysis in Theorem 1 does not hold for the sparse channel. The corresponding condition number, when $B$ is slightly larger than $B_{\rm min}$, remains very large. Thus, the proposed multi-LS approach is inadequate for the sparse channel. However, both in the considered channel and in the Rayleigh fading channel, there exists a notable phase transition around the necessary condition $B_{\rm min}=2$, resulting in a significant improvement in the condition number. This observation supports Theorem 1; nonetheless, the condition numbers at this necessary condition are not close to zero, which may lead to considerable estimation errors in practice SNR, as indicated in \eqref{eq:estimationerror}. To ensure well-conditioned scenarios, it is advisable to select the number of subframes such that  $B = 2B_{\rm min} = 4$ or $B = 3B_{\rm min} = 6$, while also satisfying the condition $QB\ll M$ (i.e., $64 \ll 512$ or $96 \ll 512$). Therefore, we conclude that the proposed multi-LS method can significantly reduce the required pilot overhead by harnessing the high rank of the considered channel model.


\subsection{Computational Complexity}\label{subsec:computationalanalysis}
In the proposed small-timescale channel estimation method, the primary computational complexity arises from the matrix inversion required to derive the multi-LS solution in \eqref{eq:LSsol}. Accordingly, the computational complexity of the proposed method is expressed as $\Oc(M^3/Q^2)$. Table~II provides a comparison of the computational complexities of the proposed and benchmark methods, where the 2TCE-FD method \cite{Hu2021} is specified in Remark 1, and the 2TCE-CLRA represents the special case of the 2TCE-PWCLRA with $Q=1$. The complexities of the 2TCE benchmark methods, based on the CLRA algorithm, depend on the estimated rank of the RIS-BS channel, where $\hat{r}_q$ and $\hat{r}$ denote the estimated rank of $\Hm_q^{\rm RB}$ and $\Hm^{\rm RB}$, respectively. In RIS-assisted near-field communications, the RIS-BS channel exhibits a high-rank, which leads to an increase in both $\hat{r}_q$ and $\hat{r}$. Nevertheless, these complexities are lower than that of the 2TCE-FD method and are comparable to the proposed 2TCE-TSP method when $\Hm^{\rm RB}$ becomes a full-rank matrix. Thus, it is asserted that the complexities of the CLRA-based method and the proposed method are nearly identical, whereas the complexity of the 2TCE-FD is regarded as having the highest computational complexity. 

To facilitate this comparison, Table III presents the average run time for deriving the LS solutions in the proposed method as a function of the number of RIS elements $M$ and the number of pieces $Q$. It is observed that the run time increases as $M$ grows and as $Q$ decreases. Particularly, when $M$ is considerably large, the computational complexity tends to decrease rapidly as $Q$ increases. In the worst case scenario of $Q=1$, it is noteworthy that the computational complexity of our method matches that of 2TCE-FD in \cite{Hu2021}. By choosing $Q > 1$, we can significantly reduce the complexity of the proposed method. Moreover, this choice is justified as it alleviates the challenges posed by the high-rank characteristic of $\Hm^{\rm RB}$. In conclusion, the proposed method, which employs piecewise estimation (i.e., $Q>1$), effectively addresses the complexity issue arising from a large number of RIS elements while preserving estimation accuracy. 

\begin{table}[t]
\renewcommand{\arraystretch}{1.5}
{\caption{Comparisons of minimum pilot overheads and computational complexities per each time block $t\geq 1$}} 
\centering

\begin{tabular}{ c|| c ||c } 
\Xhline{2\arrayrulewidth}
{Methods} & {Minimum Pilot Overheads}  & { Complexities}\\
\Xhline{2\arrayrulewidth}
{2TCE-TSP} & $QB_{\min} \approx M/N_{\rm RF}$   & $\Oc(M^3/Q^2)$\\ 
\hline
{2TCE-FD} & ${M}/\min(N_{\rm RF},{\mbox{rank}(\Hm^{\rm RB})})$ &  $\Oc\left(M^3\right)$ \\ 
\hline
{2TCE-PWCLRA} & $M$ &  $\Oc\left(\sum_{q=1}^{Q}\hat{r}_q^3\right)$ \\ 
\hline
{2TCE-CLRA} & $\lceil\mbox{rank}(\Hm^{\rm RB})/{N_{\rm RF}}\rceil{M}$ & $\Oc\left(\hat{r}^3\right)$ \\ 
\Xhline{2\arrayrulewidth}
\end{tabular}
\end{table}

\begin{remark}
    As the most relevant work, the 2TCE method referred to as 2TCE-FD, was proposed in \cite{Hu2021}, wherein $\Hm^{\rm RB}$ is estimated using dual-link pilot transmission for large-timescale channel estimation, under the assumption that the BS operates in full-duplex mode \cite{ahmed2015all}. Specifically, the BS transmits pilots to the RIS via the downlink channel using a single transmit antenna, after which the RIS reflects these pilot signals back to the BS through the uplink channel. Simultaneously, the BS receives pilots using its remaining antennas, enabling it to estimate $\Hm^{\rm RB}$ from the reflected signals. The subsequent effective channels are recovered by estimating only the $M\times 1$ vector $\hv_{t}^{\rm UR}$ for $t \in [T-1]$ as part of small-timescale channel estimation, given that $\Hm^{\rm RB}$ remains unchanged. Assuming that BS is equipped with a fully-digital beamforming structure (i.e., $N_{\rm RF}=N$), the minimum pilot overhead required for large-timescale channel estimation amounts to $2M$. Furthermore, it is evident that when employing a hybrid beamforming architecture, the minimum pilot overhead increases by a factor of $N/N_{\rm RF}$.  To optimally utilize the full-duplex system at the BS, the self-interference issue must be addressed using existing interference suppression methods \cite{ahmed2015all, zhu2015physical, masmoudi2016channel}, which incur additional computational costs. 
\end{remark}


\begin{table}[ht]
\caption{The average run time (sec) for deriving the LS solutions on the number of RIS elements and the number of pieces.}
\setlength{\tabcolsep}{5pt}
\renewcommand{\arraystretch}{1.5}

 \centering

 \begin{tabular}
  {P{30pt}|P{35pt}|P{35pt}|P{35pt}|P{45pt}}
  \toprule[1.5pt]
   \hline
   & $M=128$ & $M=256$ & $M=512$ & $M=1,024$\\ 
\hline
 $Q=1$ & 0.0026 & 0.0078 & 0.0305 & 0.1171  \\ 

 $Q=2$ & 0.0005 &0.0043 & 0.0137 &0.0613 \\

  $Q=4$ & 0.0001 & 0.0009 &0.0114 &0.0251  \\
  $Q=8$& 0.0001 & 0.0003 &0.0018 &0.0110 \\

  $Q=16$ & 0.0001 & 0.0002 & 0.0006& 0.0037 \\
\hline
 \bottomrule[1.5pt]
\end{tabular}
\end{table}

\subsection{Pilot Overhead}

In this analysis, we investigate the pilot overhead associated with the proposed 2TCE-TSP method. For the parameters $N_{\rm RF}$ and $\mbox{Rank}(\Hm^{\rm RB})$, the pilot overhead, as defined in \eqref{eq:PO}, can be approximated as follows:
\begin{equation}
    QB_{\rm min} \approx \frac{M}{\min\{N_{\rm RF}, \hat{r}_q\}}.
\end{equation} By selecting a piecewise level $Q$ such that $\hat{r}_{q} \geq N_{\rm RF}$, the pilot overhead simplifies to $M/N_{\rm RF}$, becoming independent of $Q$. Under this selection, the overall pilot overhead during the $T$ time blocks is computed as $ Q(N/N_{\rm RF}) + M + (T-1)(M/N_{\rm RF})$.
To minimize the overall pilot overhead, it is advisable to select $Q$ as the minimum value that satisfies 
$\hat{r}_{q} \geq N_{\rm RF}$. However, increasing $Q$ can reduce the computational complexity of the proposed method, defined as $\Oc(M^3/Q^2)$. Therefore, it is imperative to choose $Q$ by considering the tradeoff between overall pilot overhead and computational complexity.

\section{Simulation Results}\label{sec:sim}
In our simulations, we set the parameters as $N=128$, $M=512$, and $N_{\rm RF}=16$. For the VR matrix and vector, i.e., $\Fm$ and $\fv$, we consider i.i.d blocking effect of about $5\%$, i.e., $p=0.95$. The BS and RIS are positioned at coordinates $(100, -5, 0){\rm m}$ and $(0, 0, 5){\rm m}$, respectively, within a 3-D Cartesian coordinate system. Both the BS and RIS employ
uniform linear arrays (ULA) oriented vertically in the $x-y$ plane. The user is located at $(-d^{\rm RU}, -10, -5){\rm m}$, where $d^{\rm RU}$ is randomly selected from a uniform distribution within the range  $[20,30]{\rm m}$. We choose eight NLoS paths for both the BS-RIS and the RIS-User channels, i.e., $L^{\rm RB}=L^{\rm UR}=8$, with each path experiencing an average attenuation of $-15$ dB compared to the LoS path. The scatter locations are randomly determined, taking into account the positions of the BS, RIS, and user. 

To measure the accuracy of channel estimation, we adopt the normalized mean square error (NMSE), as outlined in related works \cite{chen2023channel,schroeder2022channel,Yang2023,Yang2024,Lee2024near,Lee2025}:
\begin{equation}
    \mbox{NMSE}\eqdef\EE\left[\frac{1}{T}\sum_{t=1}^{T}\frac{{\left\|\hat{\Hm}_{t}^{\rm eff} - \Hm_t^{\rm eff}\right\|^2}}{{\left\|\Hm_t^{\rm eff}\right\|^2}}\right],
\end{equation} where the estimated effective channels from the proposed method are expressed as follows: 
\begin{equation}
    \hat{\Hm}_t^{\rm eff} = \begin{bmatrix}
        \hat{\Hm}_{[1,0]}^{\rm pw}\mbox{diag}(\dv_{[1,t]}^{\rm LS}) &\cdots& \hat{\Hm}_{[Q,0]}^{\rm pw}\mbox{diag}(\dv_{[Q,t]}^{\rm LS})
    \end{bmatrix},
\end{equation} 
for $t \in [T-1]$. We evaluate the expectation through Monte Carlo simulations with $10^3$ trials, where the RIS-BS channel remains fixed while the User-RIS channel varies independently across trials. This methodology enables us to assess the average performance of the proposed method for rapid small-timescale channel estimation. The signal-to-noise ratio ($\SNR$) is defined as: 
\begin{equation}
    \SNR = 10\log_{10}\left[\frac{1}{T}\sum_{t=1}^{T}\frac{\left\|\Wm^{\rm RF}\Hm^{\rm RB}\mbox{diag}(\nuv_{[b,i]})\hv_t^{\rm UR}s\right\|_2^2}{\left\|\Wm^{\rm RB}\nv_{[b,i,t]}\right\|_2^2}\right].
\end{equation} 

Before discussing the simulation results, it is important to note that popular codebook-based channel estimation methods \cite{chen2023channel,Yang2023,Yang2024} will not serve as benchmarks in this analysis. As outlined in \cite{Lee2024near,Lee2025}, designing suitable codebooks for the near-field effective channel presented in Section~\ref{subsec:CM} poses significant challenges. Furthermore, it has been demonstrated that low-rank approximation-based methods, such as CLRA \cite{Lee2024near} and PW-CLRA \cite{Lee2025}, significantly outperform these codebook-based methods, such as S-MJCE \cite{chen2023channel} and 3D-M-LAOMP \cite{Yang2023}, while requiring lower pilot overhead. In our simulations, we will compare the proposed channel estimation method with the benchmarks in Table II. Lastly, we assume a perfect estimation of the initial large-time channel unless otherwise specified.

\begin{remark}
    The proposed channel estimation method, referred to as 2TCE-TSP, is applicable to various uniform antenna arrays that possess half-wavelength antenna spacing, including uniform planar arrays (UPA) and uniform circular arrays (UCA), without requiring any modifications. It is noteworthy that both the MIMO-ARD and the MIMO-RD, as defined in \eqref{eq:ARD} and \eqref{eq:RD}, are determined by the antenna apertures of the BS and the RIS. Specifically, the antenna apertures for UPA and UCA are defined by their diagonal length and diameter, respectively, while for ULA, the aperture is characterized by the length of the array. Given that the MIMO-ARD and MIMO-RD in the ULA scenario are the longest among the uniform arrays for a given number of antennas, it is essential to consider the near-field effect for practical distances, such as hundreds of meters. This consideration underpins the rationale for employing the ULA in our simulations.
\end{remark}

\begin{table}[t]
\renewcommand{\arraystretch}{1.5}
\caption{The NMSE of the benchmark 2TCE-PWCLRA \& 2TCE-CLRA on SNR for the considered near-field channel model.} 
\centering

\begin{tabular}{ c|| c | c | c  } 
\Xhline{2\arrayrulewidth}
{Methods} &  $\SNR=10$ dB  & $\SNR=20$ dB &  $\SNR=30$ dB \\
\Xhline{2\arrayrulewidth}
{2TCE-PWCLRA} & -16.0206 dB  & -26.0206 dB &  -36.0246 dB  \\ 
\hline
{2TCE-CLRA} & -13.0715 dB & -14.2366 dB & -14.3533 dB \\ 
\Xhline{2\arrayrulewidth}
\end{tabular}
\end{table}

Fig. 5 depicts the NMSE as a function of pilot overhead. It is evident that the estimation accuracy of the proposed 2TCE-TSP improves as the pilot overhead increases, regardless of the channel models, which are indicated as ``Sparse Channel \cite{Yang2024}'', ``Considered Near-Field Channel'', and ``Rayleigh Fading Channel'' in the legend. This observation suggests that by leveraging the proposed time-scaling property, the proposed method can effectively estimate the channel with a pilot overhead less than $QB = M = 512$, which represents the minimum pilot overhead required by the SOTA methods, such as 2TCE-PWCLRA and 2TCE-CLRA. When compared to the estimation accuracy of the 2TCE-PWCLRA presented in Table IV, the 2TCE-TSP achieves an overhead reduction of approximately $38\%$ and $78\%$ for the considered near-field and Rayleigh fading channels, respectively. Furthermore, the estimation accuracies of the proposed method across all channel models are comparable to or higher than those of the benchmark 2TCE-FD. These results underscore the advantages of the proposed method for the 2TCE-FD in two significant ways: i) As noted in Remark 2, the proposed method can be employed without the necessity of full-duplex mode in RIS-aided communication systems, and ii) By utilizing the proposed piecewise beam training, the proposed method substantially reduces computational complexity, as discussed in Section~\ref{subsec:computationalanalysis}.

In contrast, the proposed method struggles to estimate the effective channel in low or intermediate overhead for the sparse channel. However, it exhibits significant estimation accuracy with lower overhead for both the considered near-field and Rayleigh fading channels. This disparity arises because the condition number of the Gram matrices, denoted as $\{\Gm_q:q\in[Q]\}$, significantly improves for the considered near-field and Rayleigh fading channels even at low overhead levels. Whereas, the condition number for the sparse channel remains considerably high, as illustrated in Fig. 4. Consequently, the estimation errors articulated in \eqref{eq:estimationerror} for the considered near-field and Rayleigh fading channels diminish rapidly due to the improved condition of the Gram matrices, whereas the errors for the sparse channel persist at elevated levels. For the considered near-field channel, these findings verify our theoretical analysis of the multi-LS problem in Section~\ref{subsec:theoanalysis}. By judiciously selecting the number of subframes to slightly exceed the necessary condition, such as $QB = 16\times 3= 48$ or $QB = 16\times 4= 64$, the proposed method achieves a satisfactory level of channel estimation accuracy.

\begin{figure}[t]
\centering
\includegraphics[width=0.90\linewidth]{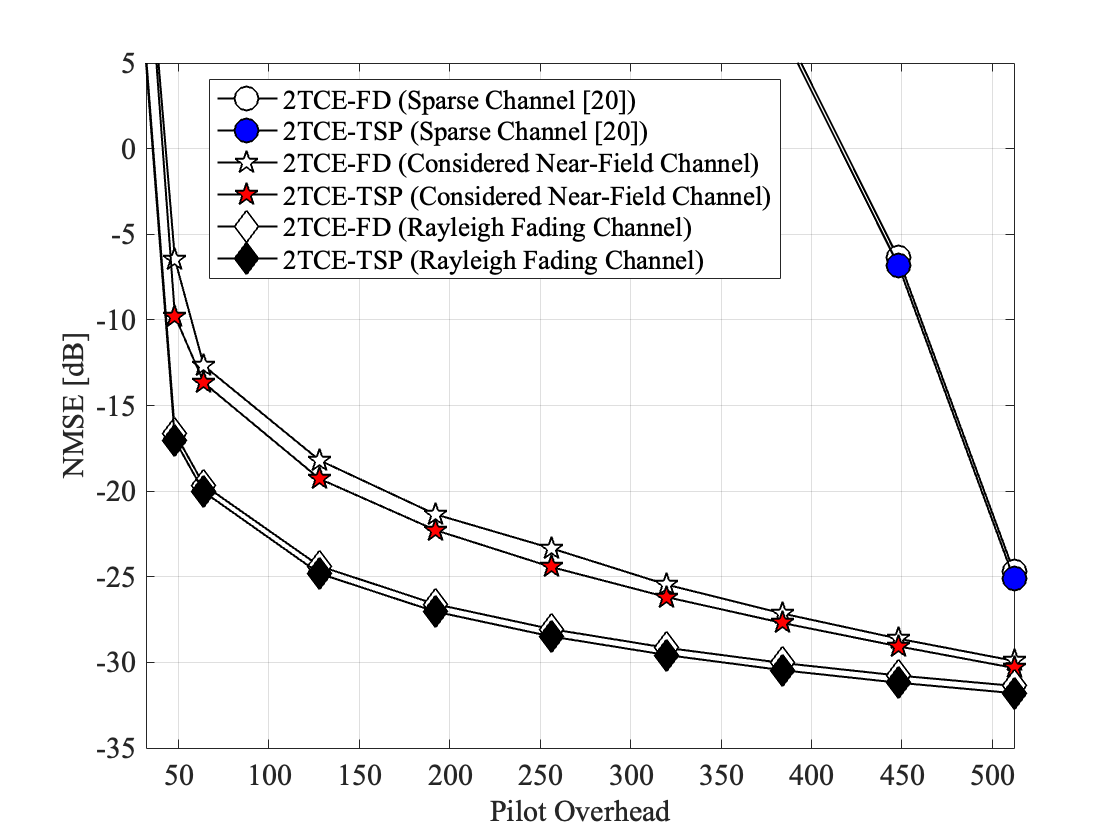}
\caption{The NMSE on the number of pilot overhead, where $Q=16$, and $\SNR=20$ dB.}
\end{figure}

\begin{figure}[t]
\centering
\includegraphics[width=0.90\linewidth]{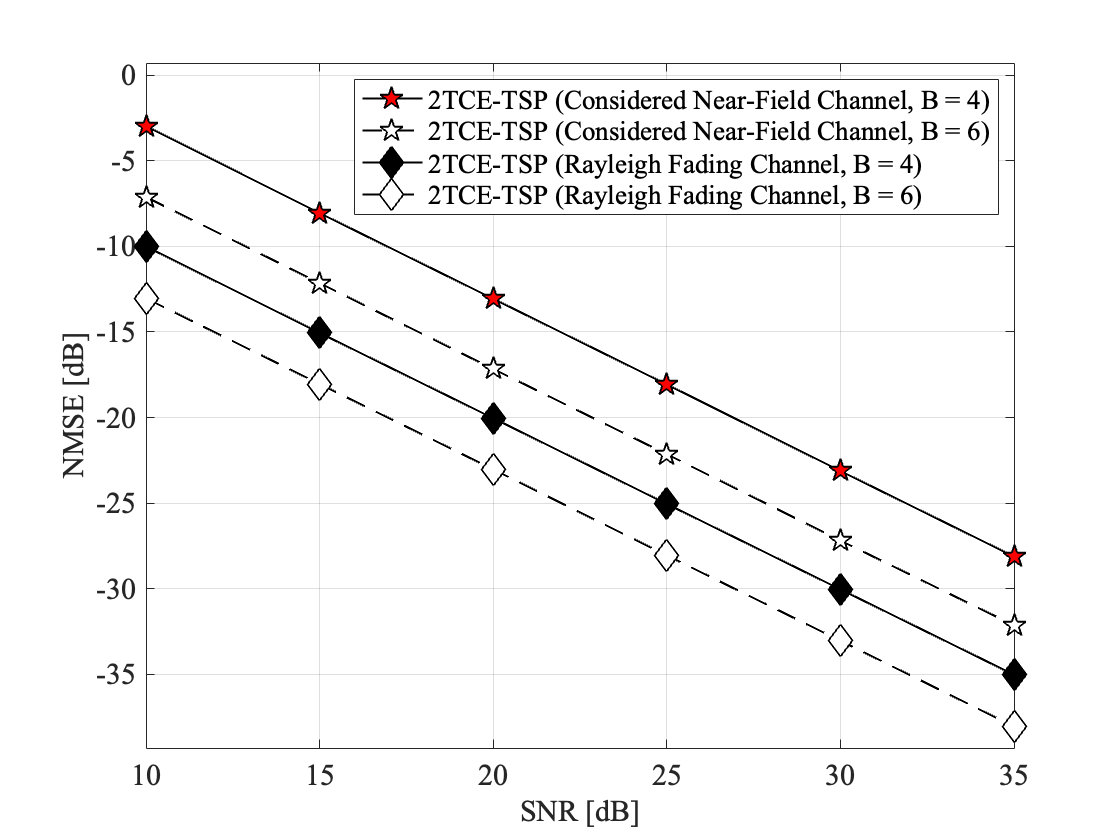}
\caption{The NMSE on SNR, where $Q=16$.}
\end{figure}

\begin{figure}[t]
\centering
\includegraphics[width=0.90\linewidth]{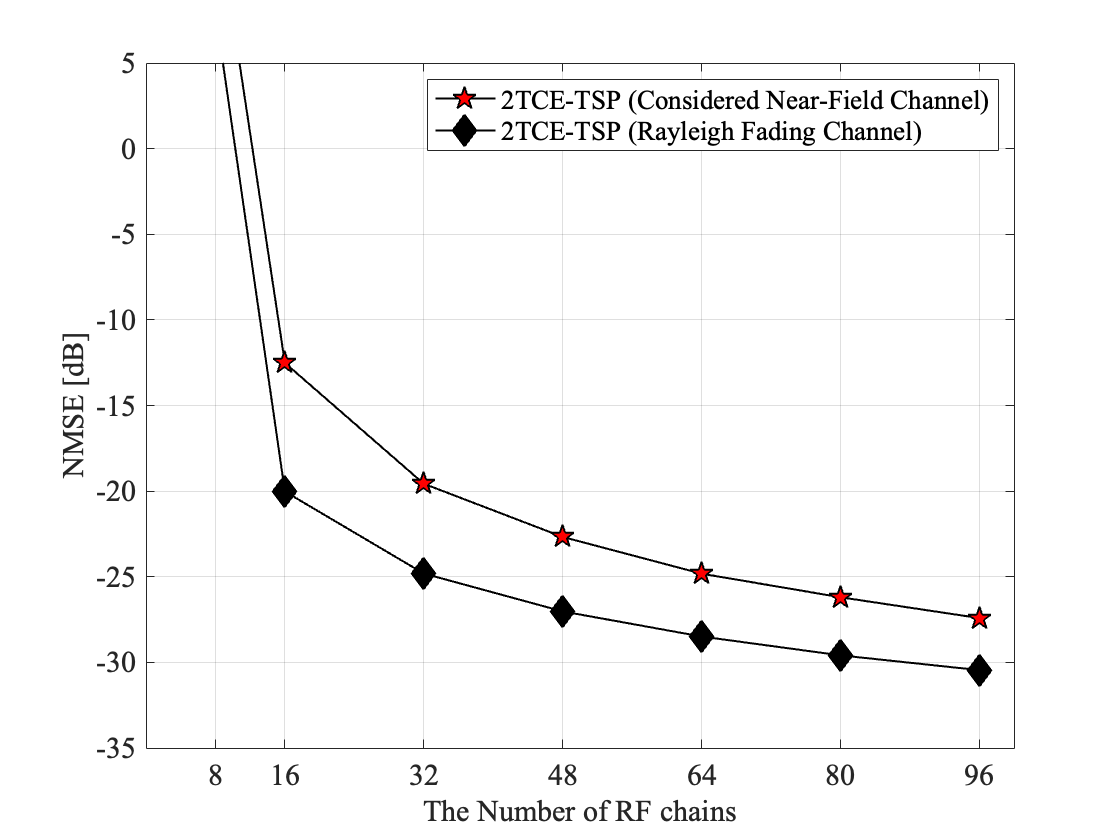}
\caption{The NMSE on the number of RF chains, where $\SNR=20$ dB, $Q=16$, and $B=4$.}
\end{figure}

\begin{figure}[t]
\centering
\includegraphics[width=0.90\linewidth]{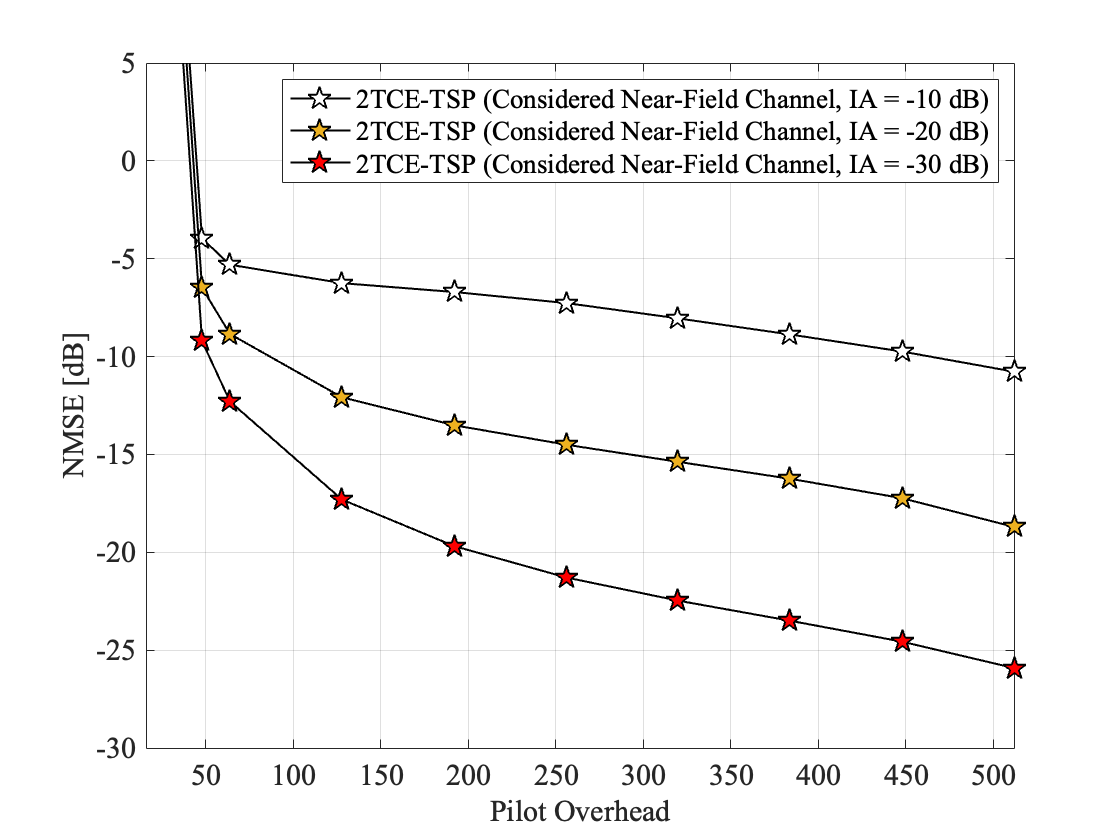}
\caption{The NMSE on pilot overhead with the initial channel estimation error, where $Q=16$ and $\SNR = 20$ dB.}
\end{figure}


Fig. 6 shows the NMSE as a function of SNRs for a low pilot overhead, specifically $QB = 64$. It is evident that the estimation accuracy of the proposed method improves as the SNR increases. By integrating the results from Fig. 5, one can determine the appropriate pilot overhead given a specific transmit power. For instance, to achieve a target accuracy of $10^{-2}$ at $\SNR=20$ dB, a minimum pilot overhead of $QB = 16\times 8 = 128$ is necessary. Conversely, as the SNR rises to $30$dB, the same accuracy can be attained with a reduced pilot overhead of $QB < 16\times 4 = 64$. This indicates that higher SNRs enable the achievement of target accuracy with significantly lower pilot overhead. Taking into account the tradeoff between total power consumption and pilot overhead, it becomes feasible to identify an optimal resource allocation for efficient communication.


Fig. 7 depicts the impact of the number of RF chains on the NMSE. It is observed that, for a given pilot overhead, the estimation accuracy improves as the number of RF chains increases. This enhancement is attributable to the equation presented in \eqref{eq:Gm}, which indicates that the condition number of $\Gm_q$ is enhanced as the column dimension of $\Am_{[b,q]}$ in \eqref{eq:Ab} expands. This phenomenon is analogous to the performance of the MIMO zero-forcing receiver \cite{Tse}, where an increased number of antennas leads to enhanced detection performance. In light of these results, it is evident that the system parameters—including transmit power, the number of pieces, and the number of RF chains—should be designed with consideration for the characteristics of each channel model. This strategic approach will ensure optimized performance in varying operational conditions.


Fig. 8 illustrates the NMSE on the pilot overhead in the presence of initial channel estimation errors, where ``IA" in the legend represents the accuracy of the initial large-timescale channel estimation. It is evident that the estimation accuracy of the proposed method in subsequent time blocks significantly improves as the IA enhances. To effectively leverage the proposed 2TCE-TSP method, it is crucial to obtain an accurate estimation of the large-timescale channel during the initial time block. It is important to note that the benchmark 2TCE-FD method requires full-duplex communication to estimate the large-timescale channel, which incurs substantial pilot overhead and computational complexity. In contrast, the proposed method capitalizes on time-scaling properties, thereby facilitating the 2TCE strategy with reduced overhead and complexity. This is particularly advantageous in practical scenarios, such as TDD with half-duplexing.

Our experimental analysis suggests that the proposed 2TCE-TSP would be a good candidate for RIS-aided near-field communication systems owing to its attractive performance, lower overhead and complexity, and practicality.

\section{Conclusion}

We investigated the channel estimation problem for RIS-aided near-field communication systems. To effectively tackle the challenges of high pilot overhead and computational complexity associated with channel estimation, we implemented a two-timescale channel estimation strategy. This strategy leverages the time-scaling property and comprises both large-timescale and small-timescale channel estimations. Utilizing the  PW-CLRA method for large-timescale channel estimation, we formulate the multiple least-squares (multi-LS) problem. This problem aims to estimate the small-timescale effective channels by employing the previously estimated large-timescale channel alongside observations gathered through the proposed beam training method. Based on our theoretical analysis, we established the efficacy of the proposed beam training method and estimated the performance of our channel estimation method. Through simulations, we validated our theoretical analysis and demonstrated notable performances of the proposed method across various real-world channels.


\begin{thebibliography}{10}
\providecommand{\url}[1]{#1}
\csname url@samestyle\endcsname
\providecommand{\newblock}{\relax}
\providecommand{\bibinfo}[2]{#2}
\providecommand{\BIBentrySTDinterwordspacing}{\spaceskip=0pt\relax}
\providecommand{\BIBentryALTinterwordstretchfactor}{4}
\providecommand{\BIBentryALTinterwordspacing}{\spaceskip=\fontdimen2\font plus
\BIBentryALTinterwordstretchfactor\fontdimen3\font minus \fontdimen4\font\relax}
\providecommand{\BIBforeignlanguage}[2]{{%
\expandafter\ifx\csname l@#1\endcsname\relax
\typeout{** WARNING: IEEEtran.bst: No hyphenation pattern has been}%
\typeout{** loaded for the language `#1'. Using the pattern for}%
\typeout{** the default language instead.}%
\else
\language=\csname l@#1\endcsname
\fi
#2}}
\providecommand{\BIBdecl}{\relax}
\BIBdecl

\bibitem{di2020smart}
M.~Di~Renzo, A.~Zappone, M.~Debbah, M.-S. Alouini, C.~Yuen, J.~De~Rosny, and S.~Tretyakov, ``Smart radio environments empowered by reconfigurable intelligent surfaces: How it works, state of research, and the road ahead,'' \emph{IEEE journal on selected areas in communications}, vol.~38, no.~11, pp. 2450--2525, 2020.

\bibitem{pei2021ris}
X.~Pei, H.~Yin, L.~Tan, L.~Cao, Z.~Li, K.~Wang, K.~Zhang, and E.~Bj{\"o}rnson, ``Ris-aided wireless communications: Prototyping, adaptive beamforming, and indoor/outdoor field trials,'' \emph{IEEE Transactions on Communications}, vol.~69, no.~12, pp. 8627--8640, 2021.

\bibitem{long2020active}
R.~Long, Y.-C. Liang, Y.~Pei, and E.~G. Larsson, ``Active intelligent reflecting surface for simo communications,'' in \emph{IEEE Global Communications Conference}.\hskip 1em plus 0.5em minus 0.4em\relax IEEE, 2020, pp. 1--6.

\bibitem{zhou2020robust}
G.~Zhou, C.~Pan, H.~Ren, K.~Wang, M.~Di~Renzo, and A.~Nallanathan, ``Robust beamforming design for intelligent reflecting surface aided miso communication systems,'' \emph{IEEE Wireless Communications Letters}, vol.~9, no.~10, pp. 1658--1662, 2020.

\bibitem{zhou2021stochastic}
G.~Zhou, C.~Pan, H.~Ren, K.~Wang, M.~Elkashlan, and M.~Di~Renzo, ``Stochastic learning-based robust beamforming design for ris-aided millimeter-wave systems in the presence of random blockages,'' \emph{IEEE Transactions on Vehicular Technology}, vol.~70, no.~1, pp. 1057--1061, 2021.

\bibitem{wu2019intelligent}
Q.~Wu and R.~Zhang, ``Intelligent reflecting surface enhanced wireless network via joint active and passive beamforming,'' \emph{IEEE transactions on wireless communications}, vol.~18, no.~11, pp. 5394--5409, 2019.

\bibitem{win2022location}
M.~Z. Win, Z.~Wang, Z.~Liu, Y.~Shen, and A.~Conti, ``Location awareness via intelligent surfaces: A path toward holographic nln,'' \emph{IEEE Vehicular Technology Magazine}, vol.~17, no.~2, pp. 37--45, 2022.

\bibitem{wang2022location}
Z.~Wang, Z.~Liu, Y.~Shen, A.~Conti, and M.~Z. Win, ``Location awareness in beyond 5g networks via reconfigurable intelligent surfaces,'' \emph{IEEE Journal on Selected Areas in Communications}, vol.~40, no.~7, pp. 2011--2025, 2022.

\bibitem{wymeersch2020radio}
H.~Wymeersch, J.~He, B.~Denis, A.~Clemente, and M.~Juntti, ``Radio localization and mapping with reconfigurable intelligent surfaces: Challenges, opportunities, and research directions,'' \emph{IEEE Vehicular Technology Magazine}, vol.~15, no.~4, pp. 52--61, 2020.

\bibitem{tsai2018efficient}
C.-R. Tsai, Y.-H. Liu, and A.-Y. Wu, ``Efficient compressive channel estimation for millimeter-wave large-scale antenna systems,'' \emph{IEEE Transactions on Signal Processing}, vol.~66, no.~9, pp. 2414--2428, 2018.

\bibitem{chen2023channel}
J.~Chen, Y.-C. Liang, H.~V. Cheng, and W.~Yu, ``Channel estimation for reconfigurable intelligent surface aided multi-user mmwave mimo systems,'' \emph{IEEE Transactions on Wireless Communications}, no.~10, pp. 6853--6869, 2023.

\bibitem{chi2011sensitivity}
Y.~Chi, L.~L. Scharf, A.~Pezeshki, and A.~R. Calderbank, ``Sensitivity to basis mismatch in compressed sensing,'' \emph{IEEE Transactions on Signal Processing}, vol.~59, no.~5, pp. 2182--2195, 2011.

\bibitem{schroeder2022channel}
R.~Schroeder, J.~He, and M.~Juntti, ``Channel estimation for hybrid ris aided mimo communications via atomic norm minimization,'' in \emph{2022 IEEE International Conference on Communications Workshops (ICC Workshops)}.\hskip 1em plus 0.5em minus 0.4em\relax IEEE, 2022, pp. 1219--1224.

\bibitem{chung2023location}
H.~Chung and S.~Kim, ``Location-aware beam training and multi-dimensional anm-based channel estimation for ris-aided mmwave systems,'' \emph{IEEE Transactions on Wireless Communications}, vol.~23, no.~1, pp. 652,666, 2023.

\bibitem{wei2021channel}
L.~Wei, C.~Huang, G.~C. Alexandropoulos, C.~Yuen, Z.~Zhang, and M.~Debbah, ``Channel estimation for ris-empowered multi-user miso wireless communications,'' \emph{IEEE Transactions on Communications}, vol.~69, no.~6, pp. 4144--4157, 2021.

\bibitem{guo2022efficient}
Y.~Guo, P.~Sun, Z.~Yuan, C.~Huang, Q.~Guo, Z.~Wang, and C.~Yuen, ``Efficient channel estimation for ris-aided mimo communications with unitary approximate message passing,'' \emph{IEEE Transactions on Wireless Communications}, vol.~22, no.~2, pp. 1403--1416, 2022.

\bibitem{lin2016terahertz}
C.~Lin and G.~Y.~L. Li, ``Terahertz communications: An array-of-subarrays solution,'' \emph{IEEE Communications Magazine}, vol.~54, no.~12, pp. 124--131, 2016.

\bibitem{molisch2017hybrid}
A.~F. Molisch, V.~V. Ratnam, S.~Han, Z.~Li, S.~L.~H. Nguyen, L.~Li, and K.~Haneda, ``Hybrid beamforming for massive mimo: A survey,'' \emph{IEEE Communications magazine}, vol.~55, no.~9, pp. 134--141, 2017.

\bibitem{wang2020joint}
P.~Wang, J.~Fang, L.~Dai, and H.~Li, ``Joint transceiver and large intelligent surface design for massive mimo mmwave systems,'' \emph{IEEE transactions on wireless communications}, vol.~20, no.~2, pp. 1052--1064, 2020.

\bibitem{yildirim2022ris}
I.~Yildirim, A.~Koc, E.~Basar, and T.~Le-Ngoc, ``Ris-aided angular-based hybrid beamforming design in mmwave massive mimo systems,'' in \emph{GLOBECOM 2022-2022 IEEE Global Communications Conference}.\hskip 1em plus 0.5em minus 0.4em\relax IEEE, 2022, pp. 5267--5272.

\bibitem{9743307}
S.~H. Hong, J.~Park, S.-J. Kim, and J.~Choi, ``Hybrid beamforming for intelligent reflecting surface aided millimeter wave mimo systems,'' \emph{IEEE Transactions on Wireless Communications}, vol.~21, no.~9, pp. 7343--7357, 2022.

\bibitem{chung2022efficient}
H.~Chung and S.~Kim, ``Efficient two-stage beam training and channel estimation for ris-aided mmwave systems via fast alternating least squares,'' in \emph{ICASSP 2022-2022 IEEE International Conference on Acoustics, Speech and Signal Processing (ICASSP)}.\hskip 1em plus 0.5em minus 0.4em\relax IEEE, 2022, pp. 5188--5192.

\bibitem{JJLee2023}
J.~Lee and S.~Hong, ``Channel estimation for ris-aided mmwave mu-mimo systems: Collaborative low-rank matrix completion approach,'' in \emph{IEEE International conference on Network Intelligence and Digital Content (IC-NIDC)}.\hskip 1em plus 0.5em minus 0.4em\relax IEEE, 2024, pp. 1--5.

\bibitem{hastie2015matrix}
T.~Hastie, R.~Mazumder, J.~D. Lee, and R.~Zadeh, ``Matrix completion and low-rank svd via fast alternating least squares,'' \emph{The Journal of Machine Learning Research}, vol.~16, no.~1, pp. 3367--3402, 2015.

\bibitem{ahmed2018survey}
I.~Ahmed, H.~Khammari, A.~Shahid, A.~Musa, K.~S. Kim, E.~De~Poorter, and I.~Moerman, ``A survey on hybrid beamforming techniques in 5g: Architecture and system model perspectives,'' \emph{IEEE Communications Surveys \& Tutorials}, vol.~20, no.~4, pp. 3060--3097, 2018.

\bibitem{ma2021joint}
X.~Ma, S.~Guo, H.~Zhang, Y.~Fang, and D.~Yuan, ``Joint beamforming and reflecting design in reconfigurable intelligent surface-aided multi-user communication systems,'' \emph{IEEE Transactions on Wireless Communications}, vol.~20, no.~5, pp. 3269--3283, 2021.

\bibitem{liu2020matrix}
H.~Liu, X.~Yuan, and Y.-J.~A. Zhang, ``Matrix-calibration-based cascaded channel estimation for reconfigurable intelligent surface assisted multiuser mimo,'' \emph{IEEE Journal on Selected Areas in Communications}, vol.~38, no.~11, pp. 2621--2636, 2020.

\bibitem{wax1985detection}
M.~Wax and T.~Kailath, ``Detection of signals by information theoretic criteria,'' \emph{IEEE Transactions on acoustics, speech, and signal processing}, vol.~33, no.~2, pp. 387--392, 1985.

\bibitem{chae2023column}
J.~Chae, P.~Narayanamurthy, S.~Bac, S.~M. Sharada, and U.~Mitra, ``Column-based matrix approximation with quasi-polynomial structure,'' in \emph{ICASSP 2023-2023 IEEE International Conference on Acoustics, Speech and Signal Processing (ICASSP)}.\hskip 1em plus 0.5em minus 0.4em\relax IEEE, 2023, pp. 1--5.

\bibitem{eckart1936approximation}
C.~Eckart and G.~Young, ``The approximation of one matrix by another of lower rank,'' \emph{Psychometrika}, vol.~1, no.~3, pp. 211--218, 1936.

\end{thebibliography}


\begin{thebibliography}{10}
\providecommand{\url}[1]{#1}
\csname url@samestyle\endcsname
\providecommand{\newblock}{\relax}
\providecommand{\bibinfo}[2]{#2}
\providecommand{\BIBentrySTDinterwordspacing}{\spaceskip=0pt\relax}
\providecommand{\BIBentryALTinterwordstretchfactor}{4}
\providecommand{\BIBentryALTinterwordspacing}{\spaceskip=\fontdimen2\font plus
\BIBentryALTinterwordstretchfactor\fontdimen3\font minus \fontdimen4\font\relax}
\providecommand{\BIBforeignlanguage}[2]{{%
\expandafter\ifx\csname l@#1\endcsname\relax
\typeout{** WARNING: IEEEtran.bst: No hyphenation pattern has been}%
\typeout{** loaded for the language `#1'. Using the pattern for}%
\typeout{** the default language instead.}%
\else
\language=\csname l@#1\endcsname
\fi
#2}}
\providecommand{\BIBdecl}{\relax}
\BIBdecl

\bibitem{dang2020should}
S.~Dang, O.~Amin, B.~Shihada, and M.-S. Alouini, ``What should 6g be?'' \emph{Nature Electronics}, vol.~3, no.~1, pp. 20--29, 2020.

\bibitem{Cha2023}
M.~Chafii, L.~Bariah, S.~Muhaidat, and M.~Debbah, ``Twelve scientific challenges for 6g: Rethinking the foundations of communications theory,'' \emph{IEEE Communications Surveys \& Tutorials}, vol.~25, no.~2, pp. 868--904, 2023.

\bibitem{Wang2024}
C.-X. Wang, X.~You, X.~Gao, X.~Zhu, Z.~Li, C.~Zhang, H.~Wang, Y.~Huang, Y.~Chen, H.~Haas, J.~S. Thompson, E.~G. Larsson, M.~D. Renzo, W.~Tong, P.~Zhu, X.~Shen, H.~V. Poor, and L.~Hanzo, ``On the road to 6g: Visions, requirements, key technologies, and testbeds,'' \emph{IEEE Communications Surveys \& Tutorials}, vol.~25, no.~2, pp. 905--974, 2023.

\bibitem{wang2018millimeter}
X.~Wang, L.~Kong, F.~Kong, F.~Qiu, M.~Xia, S.~Arnon, and G.~Chen, ``Millimeter wave communication: A comprehensive survey,'' \emph{IEEE Communications Surveys \& Tutorials}, vol.~20, no.~3, pp. 1616--1653, 2018.

\bibitem{Jiang2024}
W.~Jiang, Q.~Zhou, J.~He, M.~A. Habibi, S.~Melnyk, M.~El-Absi, B.~Han, M.~D. Renzo, H.~D. Schotten, F.-L. Luo, T.~S. El-Bawab, M.~Juntti, M.~Debbah, and V.~C.~M. Leung, ``Terahertz communications and sensing for 6g and beyond: A comprehensive review,'' \emph{IEEE Communications Surveys \& Tutorials}, vol.~26, no.~4, pp. 2326--2381, 2024.

\bibitem{Carvalho2020}
E.~D. Carvalho, A.~Ali, A.~Amiri, M.~Angjelichinoski, and R.~W. Heath, ``Non-stationarities in extra-large-scale massive mimo,'' \emph{IEEE Wireless Communications}, vol.~27, no.~4, pp. 74--80, 2020.

\bibitem{di2020smart}
M.~Di~Renzo, A.~Zappone, M.~Debbah, M.-S. Alouini, C.~Yuen, J.~De~Rosny, and S.~Tretyakov, ``Smart radio environments empowered by reconfigurable intelligent surfaces: How it works, state of research, and the road ahead,'' \emph{IEEE journal on selected areas in communications}, vol.~38, no.~11, pp. 2450--2525, 2020.

\bibitem{pei2021ris}
X.~Pei, H.~Yin, L.~Tan, L.~Cao, Z.~Li, K.~Wang, K.~Zhang, and E.~Bj{\"o}rnson, ``Ris-aided wireless communications: Prototyping, adaptive beamforming, and indoor/outdoor field trials,'' \emph{IEEE Transactions on Communications}, vol.~69, no.~12, pp. 8627--8640, 2021.

\bibitem{wu2021intelligent}
Q.~Wu, S.~Zhang, B.~Zheng, C.~You, and R.~Zhang, ``Intelligent reflecting surface-aided wireless communications: A tutorial,'' \emph{IEEE transactions on communications}, vol.~69, no.~5, pp. 3313--3351, 2021.

\bibitem{Mu2024}
X.~Mu, J.~Xu, Y.~Liu, and L.~Hanzo, ``Reconfigurable intelligent surface-aided near-field communications for 6g: Opportunities and challenges,'' \emph{IEEE Vehicular Technology Magazine}, vol.~19, no.~1, pp. 65--74, 2024.

\bibitem{Lv2024}
S.~Lv, Y.~Liu, X.~Xu, A.~Nallanathan, and A.~L. Swindlehurst, ``Ris-aided near-field mimo communications: Codebook and beam training design,'' \emph{IEEE Transactions on Wireless Communications}, vol.~23, no.~9, pp. 12\,531--12\,546, 2024.

\bibitem{Zhou2024}
H.~Zhou, M.~Erol-Kantarci, Y.~Liu, and H.~V. Poor, ``A survey on model-based, heuristic, and machine learning optimization approaches in ris-aided wireless networks,'' \emph{IEEE Communications Surveys \& Tutorials}, vol.~26, no.~2, pp. 781--823, 2024.

\bibitem{Zhou2015}
Z.~Zhou, X.~Gao, J.~Fang, and Z.~Chen, ``Spherical wave channel and analysis for large linear array in los conditions,'' in \emph{2015 IEEE Globecom Workshops (GC Wkshps)}, 2015, pp. 1--6.

\bibitem{Lu2023}
Y.~Lu and L.~Dai, ``Near-field channel estimation in mixed los/nlos environments for extremely large-scale mimo systems,'' \emph{IEEE Transactions on Communications}, vol.~71, no.~6, pp. 3694--3707, 2023.

\bibitem{Cui2024}
M.~Cui and L.~Dai, ``Near-field wideband beamforming for extremely large antenna arrays,'' \emph{IEEE Transactions on Wireless Communications}, vol.~23, no.~10, pp. 13\,110--13\,124, 2024.

\bibitem{Cui2023rainbow}
M.~Cui, L.~Dai, Z.~Wang, S.~Zhou, and N.~Ge, ``Near-field rainbow: Wideband beam training for xl-mimo,'' \emph{IEEE Transactions on Wireless Communications}, vol.~22, no.~6, pp. 3899--3912, 2023.

\bibitem{chen2023channel}
J.~Chen, Y.-C. Liang, H.~V. Cheng, and W.~Yu, ``Channel estimation for reconfigurable intelligent surface aided multi-user mmwave mimo systems,'' \emph{IEEE Transactions on Wireless Communications}, no.~10, pp. 6853--6869, 2023.

\bibitem{schroeder2022channel}
R.~Schroeder, J.~He, and M.~Juntti, ``Channel estimation for hybrid ris aided mimo communications via atomic norm minimization,'' in \emph{2022 IEEE International Conference on Communications Workshops (ICC Workshops)}.\hskip 1em plus 0.5em minus 0.4em\relax IEEE, 2022, pp. 1219--1224.

\bibitem{Yang2023}
S.~Yang, W.~Lyu, Z.~Hu, Z.~Zhang, and C.~Yuen, ``Channel estimation for near-field xl-ris-aided mmwave hybrid beamforming architectures,'' \emph{IEEE Transactions on Vehicular Technology}, vol.~72, no.~8, pp. 11\,029--11\,034, 2023.

\bibitem{Yang2024}
S.~Yang, C.~Xie, W.~Lyu, B.~Ning, Z.~Zhang, and C.~Yuen, ``Near-field channel estimation for extremely large-scale reconfigurable intelligent surface (xl-ris)-aided wideband mmwave systems,'' \emph{IEEE Journal on Selected Areas in Communications}, vol.~42, no.~6, pp. 1567--1582, 2024.

\bibitem{Lee2024near}
J.~Lee, H.~Chung, Y.~Cho, S.~Kim, and S.~Hong, ``Near-field channel estimation for xl-ris assisted multi-user xl-mimo systems: Hybrid beamforming architectures,'' \emph{IEEE Transactions on Communications}, vol.~73, no.~3, pp. 1560--1574, 2025.

\bibitem{Lee2025}
J.~Lee and S.~Hong, ``Near-field los/nlos channel estimation for ris-aided mu-mimo systems: Piece-wise low-rank approximation approach,'' \emph{IEEE Transactions on Wireless Communications}, pp. 1--1, 2025.

\bibitem{Hu2021}
C.~Hu, L.~Dai, S.~Han, and X.~Wang, ``Two-timescale channel estimation for reconfigurable intelligent surface aided wireless communications,'' \emph{IEEE Transactions on Communications}, vol.~69, no.~11, pp. 7736--7747, 2021.

\bibitem{Yu2023}
X.~Yu, W.~Shen, R.~Zhang, C.~Xing, and T.~Q.~S. Quek, ``Channel estimation for xl-ris-aided millimeter-wave systems,'' \emph{IEEE Transactions on Communications}, vol.~71, no.~9, pp. 5519--5533, 2023.

\bibitem{Lu2024}
H.~Lu, Y.~Zeng, C.~You, Y.~Han, J.~Zhang, Z.~Wang, Z.~Dong, S.~Jin, C.-X. Wang, T.~Jiang, X.~You, and R.~Zhang, ``A tutorial on near-field xl-mimo communications toward 6g,'' \emph{IEEE Communications Surveys \& Tutorials}, vol.~26, no.~4, pp. 2213--2257, 2024.

\bibitem{Priebe2013}
S.~Priebe and T.~Kurner, ``Stochastic modeling of thz indoor radio channels,'' \emph{IEEE Transactions on Wireless Communications}, vol.~12, no.~9, pp. 4445--4455, 2013.

\bibitem{chen2024}
Y.~Chen and L.~Dai, ``Non-stationary channel estimation for extremely large-scale mimo,'' \emph{IEEE Transactions on Wireless Communications}, vol.~23, no.~7, pp. 7683--7697, 2024.

\bibitem{Wu2025}
Y.~Wu, C.~Liu, Y.~Song, W.~Zhang, and Z.~Huang, ``Non-stationary channel estimation for xl-mimo with hybrid structure via low-rank matrix completion,'' \emph{IEEE Communications Letters}, pp. 1--1, 2025.

\bibitem{Palmucci2023}
S.~Palmucci, A.~Guerra, A.~Abrardo, and D.~Dardari, ``Two-timescale joint precoding design and ris optimization for user tracking in near-field mimo systems,'' \emph{IEEE Transactions on Signal Processing}, vol.~71, pp. 3067--3082, 2023.

\bibitem{Zhi2023}
K.~Zhi, C.~Pan, H.~Ren, K.~Wang, M.~Elkashlan, M.~D. Renzo, R.~Schober, H.~V. Poor, J.~Wang, and L.~Hanzo, ``Two-timescale design for reconfigurable intelligent surface-aided massive mimo systems with imperfect csi,'' \emph{IEEE Transactions on Information Theory}, vol.~69, no.~5, pp. 3001--3033, 2023.

\bibitem{ando1987singular}
T.~Ando, R.~A. Horn, and C.~R. Johnson, ``The singular values of a hadamard product: A basic inequality,'' \emph{Linear and Multilinear Algebra}, vol.~21, no.~4, pp. 345--365, 1987.

\bibitem{strang2000linear}
G.~Strang, ``Linear algebra and its applications,'' 2000.

\bibitem{ahmed2015all}
E.~Ahmed and A.~M. Eltawil, ``All-digital self-interference cancellation technique for full-duplex systems,'' \emph{IEEE Transactions on Wireless Communications}, vol.~14, no.~7, pp. 3519--3532, 2015.

\bibitem{zhu2015physical}
F.~Zhu, F.~Gao, T.~Zhang, K.~Sun, and M.~Yao, ``Physical-layer security for full duplex communications with self-interference mitigation,'' \emph{IEEE Transactions on Wireless Communications}, vol.~15, no.~1, pp. 329--340, 2015.

\bibitem{masmoudi2016channel}
A.~Masmoudi and T.~Le-Ngoc, ``Channel estimation and self-interference cancelation in full-duplex communication systems,'' \emph{IEEE Transactions on Vehicular Technology}, vol.~66, no.~1, pp. 321--334, 2016.

\bibitem{Tse}
D.~Tse and P.~Viswanath, \emph{Fundamentals of wireless communication}.\hskip 1em plus 0.5em minus 0.4em\relax Cambridge university press, 2005.

\end{thebibliography}
\end{document}